\documentclass[12pt,preprint]{aastex}

\usepackage{amsmath}

\begin{document}

\title{Formation and Evolution of Galactic Intermediate/Low-Mass X-ray Binaries}
\author{
Yong Shao$^{1,2}$ and Xiang-Dong Li$^{1,2}$}

\affil{$^{1}$Department of Astronomy, Nanjing University,
Nanjing 210093, China}

\affil{$^{2}$Key laboratory of Modern Astronomy and Astrophysics
(Nanjing University), Ministry of Education, Nanjing 210093, China}

\affil{$^{}$lixd@nju.edu.cn}

\begin{abstract}
We investigate the formation and evolutionary sequences of Galactic
intermediate- and low-mass X-ray binaries (I/LMXBs) by combining
binary population synthesis (BPS) and detailed stellar evolutionary
calculations. Using an updated BPS code we compute the evolution of
massive binaries that leads to the formation of incipient I/LMXBs,
and present their distribution in the initial donor mass vs. initial
orbital period diagram. We then follow the evolution of I/LMXBs
until the formation of binary millisecond pulsars (BMSPs). We find
that the birthrate of the I/LMXB population
is in the range of $ 9\times10^{-6} - 3.4\times10^{-5} \,
{\rm yr^{-1}}$, compatible with that of BMSPs which are
thought to descend from I/LMXBs. We show that during the evolution
of I/LMXBs they are likely to be observed as relatively compact
binaries with orbital periods $ \lesssim $ 1 day and donor masses
$\lesssim 0.3 M_{\odot}$. The resultant BMSPs have orbital periods
ranging from less than 1 day to a few hundred days. These features
are consistent with observations of LMXBs and BMSPs. We also confirm
the discrepancies between theoretical predications and observations
mentioned in the literature, that is, the theoretical average mass
transfer rates ($ \sim 10^{-10} M_{\odot} $\,yr$^{-1}$) of LMXBs are
considerably lower than observed, and the number of BMSPs with
orbital periods $\sim 0.1-10$ day is severely underestimated. These
discrepancies imply that something is missing in the modeling of
LMXBs, which is likely to be related to the mechanisms of the
orbital angular momentum loss.

\end{abstract}


\section{Introduction}

Low-mass X-ray binaries (LMXBs) consist of an accreting compact
star, either a black hole or a neutron star (NS) and a low-mass ($
\lesssim 1 M_{\odot} $) donor star, in which mass transfer proceeds
via Roche-lobe overflow (RLOF). There are about 200 LMXBs in the
Galaxy \citep{lvv07}.  The formation of black hole LMXBs is still a
controversial topic \citep[][for a review]{l15}, and will be
discussed elsewhere. Here we focus on the formation and evolution of
LMXBs with a NS. The mainstream idea is that they either have an
initially low-mass secondary, or descend from systems with initially
intermediate-mass secondaries, i.e., intermediate-mass X-ray
binaries \citep[IMXBs;][]{pr00,k00,tvs00,prp02}. Considering the
fact that NSs must have formed from massive stars, and current LMXBs
usually reside in relatively compact binary orbits \citep[with
periods $P_{\rm orb} \lesssim 10 $ days; see][]{lvv07,w10,r11}, the
progenitors of these systems should generally have experienced
common envelope (CE) evolution \citep{p76}. During the CE phase and
the subsequent supernova (SN) explosion, only a small part of the
binaries can survive as I/LMXBs \citep{prp03}. For reviews on the
formation and evolution of I/LMXBs, see \citet{bv91} and
\citet{tv06}.

The stability of the mass transfer in I/LMXBs depends on the mass
ratio, the orbital period, and  the mass and angular momentum loss
\citep[AML;][]{spv97}. Dynamically unstable mass transfer leads to
a (second) CE evolution, with the NS spiraling into the
envelope of the donor star. If the binary survives without
merging, it becomes a compact binary with a helium (He)
star around the NS. The subsequent mass transfer from the He star
may lead to the formation of a partially recycled pulsar with a CO
white dwarf (WD) companion \citep{vt84,clx11}. For those binaries
avoiding the CE phase, \citet{ps88,ps89} pointed out that there
exists a bifurcation period ($P_{\rm bif}$) for the initial orbital
period, which separates I/LMXBs into converging or diverging
systems. The existence of this bifurcation period is due to the
balance between the orbital expansion caused by nuclear
evolution of the donor star and the orbital shrinking caused by
AML. Binaries with $P_{\rm orb} < P_{\rm bif}$ continue to contract
along the cataclysmic variable (CV)-like or ultra-compact X-ray
binary (UCXB) evolutionary tracks, while those with $P_{\rm orb} >
P_{\rm bif}$ will become relatively wide binaries containing a
recycled NS and a He or CO WD
\citep{r95,kw96,kr99,ts99,tvs00,pr00,k00,l02,d08,db10,l11,jl14,itl14}.

\citet{prp02} made a systematic investigation of the evolution of
I/LMXBs by using a Henyey-type stellar evolution code to calculate
100 binary evolution sequences. In that work, the initial mass of
the NS was fixed as $ 1.4 M_{\odot} $, the initial mass of
the donor ranged from $ 0.6-7 M_{\odot} $, and the initial
orbital period went from 4 hr to 100 days. While the
calculated results showed a variety of evolutionary
channels which may explain the large diversity in observed LMXBs,
they also indicated a number of remaining problems, in
particular the low median X-ray luminosities as compared to those of
observed LMXBs.

\citet{prp03} firstly combined the binary population synthesis (BPS)
method with detailed stellar evolutionary sequences of \citet{prp02}
to follow the evolution of a population of I/LMXBs. For early Cases
B and C mass transfer\footnote{Cases A, B and C means that the donor
stars are in core hydrogen burning, shell hydrogen burning, and
after exhaustion of core helium burning stages, respectively.} they
adopted a fixed value of the critical ratio of the donor mass and
the NS mass ($ q_{\rm c} = 2 $) to determine whether the mass
transfer is stable. Late Cases B and C mass transfer was
always assumed to be dynamically unstable followed by a CE phase.
When dealing with CE evolution, they adopted constant values of the
binding energy parameter $ \lambda  = 0.1$, 0.3 and 0.5, which
parameterizes the structure of the donor's envelope.
Their calculations demonstrated that incipient IMXBs outnumber
incipient LMXBs typically by a factor of $\gtrsim 5$. It was also
found that rather large values of  $ \lambda \sim 0.5$ are required
to yield the I/LMXB birthrates that are consistent with the BMSP
birthrates, but such values of $\lambda$ would lead to an
overproduction of luminous LMXBs in the Galaxy. There were
still large discrepancies between the orbital period distributions
of modeled and observed binary millisecond pulsars (BMSPs). A
possible solution to the above problems was to reduce the
mean X-ray lifetime of LMXBs  by a factor of $\gtrsim 100$.

In this work, we revisit the formation and evolution of Galactic
I/LMXBs. Similar as \citet{prp03}, we perform a systematic
evolutionary study of I/LMXBs, combining BPS with detailed binary
evolutionary sequences. To follow the formation and evolution of
I/LMXBs, we take into account updated theoretical results and
techniques on binary evolution achieved during the last ten years.
We use the rapid binary evolution algorithm developed by \citet{h02}
for BPS calculation, which enables modeling of a large number of
complex binary systems to obtain the incipient population of
I/LMXBs, and an updated version of the \citet{e71,e72} code to
calculate the detailed evolution of individual I/LMXBs. With them we
can follow the evolutionary tracks of the I/LMXB population from the
incipient stage of a detached binary containing a newborn NS to the
RLOF phase,  and finally to the remnant state with a BMSP. The
parameter distributions of I/LMXBs and BMSPs can be clearly
displayed, and compared with observations.

The most important and difficult to understand processes during
binary evolution are unstable mass transfer and CE evolution
\citep[][for a recent review]{i13}. In the BPS calculations, instead
of using the empirical relations for the critical mass ratios, we
employ the numerically calculated results of \citet{sl14}  to judge
the stability of mass transfer in a primordial binary, and the
parameter space obtained by \citet[hereafter Paper I]{sl12b} to
determine whether I/LMXBs can successfully evolve to be binary
pulsars. We also adopt the calculated values of $ \lambda$ given by
\cite{xl10} to calculate the orbital change during the CE phase,
which depend on the evolutionary state of the donor. Other
improvement will be mentioned below. Although these updated
parameters are still subject to large uncertainties, they provide
physically more reasonable and realistic input as compared to
previous BPS studies.

This paper is organized as follows. In Section 2
we describe the methods used in this paper.
We present the detailed results derived from our calculations and
discuss their implications in Section 3. We conclude in Section 4.

\section{Methods and Calculations}

\subsection{Formation of Incipient I/LMXBs}

We use the BPS code developed by \citet{h02} and modified by
\citet{kh06} to generate the incipient I/LMXB population. This code,
based on single star evolution with a series of analytic formulae
\citep{h00}, includes standard assumptions for the population of
massive primordial binaries and analytic prescriptions to describe
binary interactions, mass transfer and SN explosions. We have
updated the code in various aspects, in particular the formation and
evolutionary processes of compact objects. Some key points are
described below \citep[see][for more details]{sl14}.

The stability of mass transfer is determined by the change of the
donor's radius with respect to its RL radius. Usually a critical
mass ratio $ q_{\rm c} $ (= donor mass/accretor mass) is used as a
criterion: if the mass ratio is larger than $ q_{\rm c} $ at the
onset of RLOF, the mass transfer is unstable and results in a CE
evolution; otherwise the mass transfer proceeds stably on a nuclear
or thermal timescale. In previous studies, the values of  $ q_{\rm
c} $ were usually estimated by only comparing the mass-radius
exponents for the donor stars, which describe the response of the
star and the RL to mass loss \citep{hw87}. However, it is well known
that mass accretion can spin up the accretor \citep{p81} and
probably result in mass loss \citep{d09}. Rapid mass accretion can
drive the accreting star out of thermal equilibrium and
cause it to expand, and this expansion may even cause the accretor
to fill its own RL, leading to the formation of a contact binary
\citep{ne01}. \citet{sl14} numerically calculated the values of $
q_{\rm c} $ for a grid of binaries with different values of
component masses and orbital periods, considering the responses of
both the donor to mass loss and the accretor to mass accretion.
Since the stellar expansion strongly depends on the accretion rate,
they constructed three models with different assumed mass accretion
histories. In Model I, the mass accretion rate depends on the
rotational velocity of the accretor. Since a small amount of
transferred material is able to spin it up to be critically rotating
\citep{p81}, there is actually very small mass accreted in this
case. In Model II, it is assumed that half of the transferred mass
is accreted and the other half is ejected out of binary system. In
model III, the mass transfer is assumed to be almost conservative.
The obtained values of $ q_{\rm c} $ vary with the orbital period,
and can reach $ \sim6 $ in Model I, but are generally less than $
\sim 2.5 $ and $ \sim 2.2 $ in Models II and III, respectively.

For unstable mass transfer, we use the standard energy conservation
equation \citep{w84} to describe the subsequent CE evolution,
\begin{equation}
\alpha_{\rm CE}(\frac{GM_{\rm 1,f}M_{2}}{2a_{\rm f}} -\frac{GM_{\rm
1,i}M_{2}}{2a_{\rm i}}) =\frac{GM_{\rm 1,i}M_{\rm 1,env}}{\lambda
R_{\rm 1,L}},
\end{equation}
where the indices 1 and 2 denote the primary and the secondary, and
i and f the initial and final values, respectively; $ M_{\rm 1,env}
$ is the mass of the primary$'$s envelope that is ejected out of the
binary during the CE phase,  $ R_{\rm 1,L} $ is the RL radius of the
primary at the onset of RLOF, and $ \alpha_{\rm CE} $ is the CE
efficiency with which the orbital energy is used to unbind the
stellar envelope. It has been pointed out that the internal energy
within the envelope can also contribute to envelope ejection, and
thus modify the values of $ \lambda$
\citep[e.g.,][]{kool90,dewi00,prh03}. We employ the  values of
$\lambda$ calculated by \citet{xl10} for a range of massive and
intermediate-mass stars at various evolutionary stages, and take $
\alpha_{\rm CE} $ = 1.0 in the calculations.

We assume that NSs are born from either core-collapse  or
electron-capture SNe, depending on the masses of the progenitor
stars. We follow the criterion suggested by \citet{b08} and
\citet{f12} to distinguish them. The He core mass at the
base of asymptotic giant branch is used to decide the formation of
various CO cores. (1) If the He core mass is less than $ 1.83
M_{\odot} $, the star forms a degenerate CO core and ends up forming
a CO WD. (2) If the core is more massive than $ 2.25 M_{\odot} $,
the star ultimately forms a FeNi core and collapses to a NS. (3)
Stars with core masses between $ 1.83 M_{\odot} $ and $ 2.25
M_{\odot} $ form partially degenerate ONe cores. If in subsequent
evolution the core mass increases to $ 1.38 M_{\odot} $, then the
core collapses to a NS due to electron capture; otherwise the star
leaves an ONe WD. Under these prescriptions, for single stars, the
stellar masses for core-collapse and electron-capture SNe are in the
range of $ 8.3-20 M_{\odot} $, and $ 7.6-8.3 M_{\odot}$,
respectively \citep{b08}. The masse interval for electron-capture
SNe is in rough accordance with \citet{wh15}. It depends on the
assumptions for mass loss and the dredge-up process in stellar
evolution \citep[e.g.,][]{s06,p08,j14}, and is blurred by mass
transfer in binaries \citep{h02}. The newborn NSs are assumed to
receive a kick due to asymmetric SN explosions, and we utilize a
Maxwellian distribution for the kick velocity,
\begin{equation}
p(v_{\rm k}) = \sqrt{\frac{2}{\pi}}\left(\frac{v_{\rm k}^{2}}{\sigma_{\rm k}^{2}}\right)
e^{-v_{\rm k}^{2}/2\sigma_{\rm k}^{2}},
\end{equation}
assuming that the directions of the kicks are isotropically distributed.
We adopt $ \sigma_{\rm k} = 265\, \rm km\, {\rm
s}^{-1} $ \citep{h05} and $ 50\, \rm km\, {\rm s}^{-1} $ \citep{d06}
for NSs formed from core-collapse and electron-capture  SNe,
respectively.

The initial parameters in our BPS calculation are taken as follows.
All the binary orbits are assumed to be circular due to tidal
interactions \citep{h02}. The mass $M_{1}$ of the primary is in the
range of $3 - 30 M_{\odot}$, following the initial mass function of
\citet{k93}\footnote{In the rapid SN mechanism of
\citet{f12}, the upper limit for the masses of the NS progenitors in
binaries is $28 M_{\sun}$.}. The initial mass ratio $M_{2}/M_{1}$
has a uniform distribution between 0 and 1, but only binaries with
the secondary mass $ M_{2} $ between $0.1\, M_{\odot}$ and $8\,
M_{\odot}$ are chosen. The logarithm of the orbital separation $a$
is uniformly distributed, lying between $3R_{\odot}$ and $10^{4}
R_{\odot}$. We adopt solar metallicity, and assume a constant star
formation rate of $5 M_{\odot}\, {\rm yr}^{-1} $ over the past 12
Gyr \citep{s78}. \citet{rw10} obtained a smaller star
formation rate of $\sim 1 M_{\odot}\, {\rm yr}^{-1} $ derived from
{\em Spitzer} detected pre-main-sequence stars. If we adopt the
latter value, the predicted number of Galactic I/LMXBs will be
reduced by several times, but the features of their distribution
will not change.

In Fig.~1, we plot the birthrates of incipient I/LMXBs as a function
of the donor mass $M_{\rm d}$ (in the range of $0.3 - 6 M_{\odot}$).
In the left and right panels the NSs are assumed to form from
core-collapse and electron-capture SNe, respectively. The green, red
and black curves correspond to the results with the $ q_{\rm c} $
values from Models I, II and III, respectively. It is clearly seen
that, in both core-collapse and electron-capture SN channels, IMXBs
dominate the whole population as expected. Although core-collapse
SN-IMXBs in Model I have a birthrate higher than in other models by
$\lesssim 50\%$, the overall birthrate distributions do not show a
remarkable difference. Thus we only take the calculated birthrates
in Model II in the following study, not only because it is
intermediate between the other two more extreme ones, but also
because the corresponding results can better match the observed
distribution of Be/X-ray binaries \citep{sl14}.

In Fig.~2 we plot the distributions of incipient I/LMXBs in
the orbital period - donor mass plane (left panel: the core-collapse
SN channel; right panel: the electron-capture SN channel). After SNe
the binary orbits become eccentric due to mass loss and the kicks
imparted on the NSs. Here we assume that they are quickly
circularized by tidal torques with the orbital AM conserved. The
corresponding orbital separation is then set to be $ a_{\rm
SN}(1-e)$, where $ a_{\rm SN} $ is the semimajor axis after the SNe.
Also shown are the histogram distributions of the birthrates, same
as the red curves in Fig.~1. In the core-collapse SN channel, the NS
progenitors masses are $8.6-28 M_{\odot}$, the resultant I/LMXBs
have orbital periods $P_{\rm orb} \lesssim 100$ days, and most of
them have $P_{\rm orb}<$ a few days. The I/LMXBs birthrate roughly
increases with the donor mass, because systems with lower-mass
companions are more likely to merge during the CE phase \citep[see
also][]{prp03}. In the electron-capture SN channel, the masses of
the NS progenitors lie in the range $7.6-8.6 M_{\odot}$. The
primordial binaries should have orbital periods longer than $\sim
1000$ days, otherwise the mass transfer starts too early so that a
WD rather a NS is produced. The binding energies for such stars are
relatively small, with the values of $ \lambda $ greater than unity
when on asymptotic giant branch \citep{dewi00,xl10}.
So the resultant I/LMXBs have orbital periods in a wide range $\sim
0.2-1000$ days, but the majority are relatively long period ($P_{\rm
orb}>10 $ days) systems. The I/LMXB birthrate slightly decreases
with increasing donor mass. The reason is that, for relatively
massive companion stars, the orbital periods after CE are usually
long because there are sufficient orbital kinetic energies. The
second mass transfer may not occur, so the primaries undergo
core-collapse SNe, which can readily disrupt the wide binaries.
Finally there are no binaries with $ P_{\rm orb} \gtrsim 1$ day and
$ M_{\rm d} \lesssim 1M_{\odot} $, because magnetic braking (MB) and
gravitational radiation (GR) cannot drive these systems into contact
within Hubble time.

\subsection{Evolution of I/LMXBs}

Based on the results in last subsection we then employ an updated
version of the \citet{e71,e72} code to calculate the evolution of
I/LMXBs. Before doing this, we first select the binaries in which
mass transfer is stable so that they can evolve to be binary
pulsars. Super-Edddinton accretion onto NSs in IMXBs may
lead to delayed dynamical instability, and eventually CE evolution.
However, there are observational and theoretical hints for mass loss
from the accreting NSs that can stabilize the mass transfer
\citep[][and references therein]{kr99,pr00}. \citet{tvs00} surveyed
the allowed parameter ($M_{\rm d}-P_{\rm orb}$) space for stable
mass transfer, and showed that the donor masses $ M_{\rm d} $ can be
up to $5 M_{\odot} $ for a $1.3M_{\sun}$ NS. Observations of
high-mass X-ray binaries, in which the NSs have accreted very small
mass, show that the initial masses of NSs range from $\sim
1\,M_{\odot}$ for 4U 1538$-$52 to $\sim 1.8\,M_{\odot}$ for Vela
X$-$1 \citep{Rawls11}. In Paper I we extended the calculation by
considering the NS masses $M_{\rm NS}$ in the range of $1.0-1.8
M_{\odot}$. The results are used to compare with the obtained
incipient I/LMXB distribution in Fig.~3 (binaries formed from both
core-collapse and electron-capture SN channels are included). The
left and right panels are for NSs with mass $1.0 M_{\odot} $ and
1.8$ M_{\odot} $, respectively. The binaries distributed in the
regions confined by the thick grey curves can stably evolve without
entering the CE phase, and are selected for further investigation.
Their corresponding birthrates are shown with the blue curves,
totally ranging from $9\times10^{-6}$ to $3.4\times10^{-5} \rm \,
yr^{-1} $. Figure 4 demonstrates the distributions of the birthrates
for these systems as a function of the initial donor masses and the
orbital periods. Here the bin size of the donor mass is 0.2 $
M_{\odot} $ and the orbital period (in units of days) changes
logarithmically in steps of 0.2.

We calculate the evolution of the selected I/LMXBs in a similar way
as in Paper I, taking into account AML caused by MB \citep{vz81,r83}
and GR \citep{ll59,f71}. The accretion rate of a NS is assumed to be
limited by the Eddington accretion rate, and we adopt the isotropic
re-emission model, assuming that the excess material leaves the
binary in the form of isotropic winds from the NS, carrying  off the
NS$'$s specific orbital AM.

Figures~5 and 6 present two example evolutionary tracks. In Fig.~5
the initial parameters are $ M_{\rm NS} = 1.8 M_{\odot}$, $ M_{\rm
d} = 1.0 M_{\odot}$ and $ P_{\rm orb}  = 1.0$ day. The left, middle,
and right panels depict the evolution of the orbital period $ P_{\rm
orb} $ with the age, the evolution of the donor radius $ R_{\rm d} $
and the mass transfer rate $ |\dot{M}_{\rm d}| $ with the donor mass
$ M_{\rm d} $, respectively. The donor starts to fill its RL and
initiate (Case A) mass transfer at $t=2.23$ Gyr when $ P_{\rm orb}=
0.33$ day, mainly driven by MB. The mass transfer rate first goes up
to $ \sim 2\times10^{-9} M_{\odot}$\,yr$^{-1}$, then decreases
gradually as the orbital period reduces to $ \sim  0.1$ day.
Meanwhile, the donor mass decreases to be less than  $ 0.3 M_{\odot}
$, and the donor becomes fully convective, so MB stops working. The
evolution afterwards is solely driven by GR, and the mass transfer
rate drops down to $ \sim 10^{-10} M_{\odot}$\,yr$^{-1}$. At the end
of the evolution (the age of $\sim 6$ Gyr), the donor mass and the
orbital period are reduced to $ \sim 0.08 M_{\odot} $ and $ \sim
0.06 $ day respectively. The duration of the mass transfer is about
3.77 Gyr.

In Fig.~6 we show the evolutionary path with the same component
masses but $ P_{\rm orb} =  6.3$ days. The mass transfer starts at
the age of 11.9 Gyr, when the donor evolves to become a subgiant.
During this Case B mass transfer, the orbital period increases all
the time, up to 73 days. The mass transfer rate varies in the range
$ 10^{-9}-10^{-8} M_{\odot}$\,yr$^{-1}$ \citep[the gap in the mass
transfer rate is caused by the so-called ``bump-related''
detachment;][]{dvl06}. The binary finally leaves a NS with a He WD
with mass of $ 0.31 M_{\odot} $. The mass transfer liftime is about
0.14 Gyr.

The above examples suggest that, given the same donor star, the
longer the initial orbital periods, the higher the averaged mass
transfer rates, and the shorter the X-ray lifetimes
\citep{w83,k88}. This strongly influences the
characteristics of observable LMXBs.

\section{Results and Discussion}

\subsection{Orbital periods, accretion rates and donor star masses of Galactic I/LMXBs}

According to the calculated evolutionary sequences of the I/LMXBs,
we plot their number distribution in the orbital period - mass
transfer rate ($ P_{\rm orb}-|\dot{M}_{\rm d}| $) plane (Fig.~7).
The left and right panels correspond to the results  with a $1.0
M_{\odot} $ and $1.8 M_{\odot} $ NS, respectively. Each panel
contains $45 \times 64$ matrix elements, in which $ P_{\rm orb} $
changes logarithmically from 0.0316 day to 1000 days with steps of
0.1, and $|\dot{M}_{\rm d}|$ from $ 10^{-12} M_{\odot}$\,yr$^{-1}$
to $10^{-4} M_{\odot}$\,yr$^{-1}$ with steps of 0.125. The color
reflects the number of I/LMXBs in the corresponding matrix element
by accumulating the product of the birthrates of X-ray binaries
passing through it with the time-span.

A comparison of Figs.~4 and 7 clearly demonstrates the change in the
orbital periods during the evolution of I/LMXBs. As we have already
seen from Figs.~5 and 6, in X-ray binaries with initially long
orbital periods, the donors have evolved to be (sub)giant stars at
the onset of mass transfer, and the high mass transfer rates imply a
short X-ray lifetime; in short-period X-ray binaries, the donors are
still on main-sequence during mass transfer, so the systems have a
long-lasting mass transfer phase. This difference in the
evolutionary time leads to different number distribution in Fig.~7,
which explains why most I/LMXBs tend to have relatively short
orbital periods ($ \lesssim 1$ day). Another factor that influences
the number distribution of LMXBs is their transient behavior. The
general idea is that, if the mass transfer rate is smaller
than a critical mass transfer rate for a given orbital period, the
accretion disk is likely subject to thermal and viscous instability,
and the LMXB may appear as a transient X-ray source \citep{l01}. In
LMXBs irradiation from the NS may help stabilize the disk to some
extent by increasing the surface temperature
\citep{v96,k97,l01,r08}. The critical mass transfer rate for the
disk instability is given by \citep{dlhc99},
\begin{equation}
\dot{M}_{\rm cr}\simeq 3.2\times10^{-9}\left(\frac{M_{\rm
NS}}{1.4M_{\odot}}\right)^{0.5} \left(\frac{M_{\rm
d}}{1.0M_{\odot}}\right)^{-0.2} \left(\frac{P_{\rm orb}}{1.0\,{\rm
d}}\right)^{1.4} M_{\odot}\, {\rm yr}^{-1},
\end{equation}
and plotted with the red dashed line. We assume that LMXBs under
this line experience short outbursts separated by long quiescent
intervals. The majority of long-period LMXBs are transient
sources spending most of their time in
quiescence\footnote{It is interesting to note that all
accreting X-ray MSPs have orbital periods less than 19 hours,
whereas radio MSPs are found up to $ P_{\rm orb}\sim200$ days. This
puzzle can be partly  understood by the transient behavior of X-ray
MSPs \citep[][and references therein]{pw12}.}. These
results are consistent with \citet{z12}, who showed that
more than $90\%$ persistent NS LMXBs have main-sequence donors and
orbital periods between $ \sim  1 - 100$ hours. Comparing the two
panels of Fig.~7 also shows that, LMXBs with a heavier NS have a
larger coverage in the parameter space, since  mass transfer in
these binaries are more likely to be stable (see Fig.~3).

In Fig.~7 we also plot Galactic LMXBs with known orbital periods and
accretion rates. Here the observational data are taken from Ritter's
LMXB catalogue
\citep{r03}\footnote{http://www.mpa-garching.mpg.de/RKcat/}, and
from \citet{r11} and \citet{c12}. The filled circles and triangles
represent persistent and transient LMXBs, respectively
\footnote{When estimating the average accretion rates for
transient LMXBs, \citet{c12} ignored the possibility of accretion in
quiescence and mass loss through winds and jets, so the estimates
should be considered as lower limits to the actual mass transfer
rates.}. It is clearly seen that most LMXBs are compact systems
except two sources: Cyg X$-$2 with a $\sim 0.6 M_{\odot}$ companion
star in a 9.84 day orbit \citep{c98,o99}, and GX13$+$1 (shown with
the star symbol), which has the longest known orbital period
(of 601.7 hours) for a Galactic NS LMXB powered by RLOF
\citep{b99}.

Figure 8 depicts the distributions of the masses (left panel) and
effective temperatures (right panel) of the donor stars. The upper
and lower panels are for persistent and transient LMXBs, and the
black and red lines correspond to $1M_{\sun}$ and $1.8M_{\sun}$ NSs,
respectively. It is seen that LMXBs dominate the population at the
current epoch, because the initial phase of thermal-timescale mass
transfer in IMXBs, where a large fraction of the secondary mass is
removed, is relatively short-lived. The donor masses in LMXBs
cluster around $ 0.2- 0.3 M_{\odot} $.  The donors' temperatures are
distributed in the range of $\sim 2500-6300$ K.

Figure 9 presents a more detailed comparison of the modeled orbital
period (left panel) and accretion rate (right panel) distributions
(in red lines) with observations (in black lines). The upper and
lower panels are for $1.0M_{\sun}$ and $1.8M_{\sun}$ NSs,
respectively. The numbers for observed sources are amplified by a
factor of 500. Although there seems broad agreement between the
measured and predicted orbital periods, there exist a systematic
discrepancy in the accretion rates, i.e., the calculated average
accretion rates ($ \sim 10^{-10} M_{\odot}$\,yr$^{-1}$) are lower
than derived from observations by about an order of magnitude.

To quantitatively compare the univariate distributions of
the calculated and observed quantities, we use Kolmogorov-Smirnov
test to assist inference on their consistency \citep{fj12}. The
cumulative distribution functions (CDFs) for the calculated and
measured orbital periods of the I/LMXBs are expressed as $ F(P_{\rm
orb}) $ and $ F_{0}(P_{\rm orb}) $, respectively.
We want to test the null hypothesis that $ F(P_{\rm orb}) =
F_{0}(P_{\rm orb})$ for all $ P_{\rm orb} $, against the alternative
that $ F(P_{\rm orb}) \neq F_{0}(P_{\rm orb})$ for some $ P_{\rm
orb} $. By comparing the CDFs, we can measure the maximum distance $
M_{\rm KS} = \sqrt{\frac{mn}{m+n}}\max |F(P_{\rm orb})-F_{0}(P_{\rm
orb})|$, where $ m $ and $ n $ are the numbers of X-ray binaries in
the calculation and the observation, respectively. Since $m\gg n$,
$M_{\rm KS} \simeq \sqrt{n}\, \max\,|F(P_{\rm orb})-F_{0}(P_{\rm
orb})|$. At a given significance level $\alpha =0.05 $, there exists
a critical value $ M_{\rm KS}^{\rm cr}\simeq 1.33$, and $ M_{\rm KS}
> M_{\rm KS}^{\rm cr}$ allows rejection of the null hypothesis,
indicating that the two distributions are significantly different.
Figure~10 shows the CDFs for the orbital period (left panel) and the
accretion rate (right panel) of the I/LMXBs. The black, red and
green curves represent the calculated CDFs with the NS mass of $1.0
M_{\odot} $ and $1.8 M_{\odot} $, and the observed CDF,
respectively. For the orbital period distribution, we obtain $
M_{\rm KS} = 2.62$ and 1.15 with a $1.0 M_{\odot} $ and $1.8
M_{\odot} $ NS, respectively; while for the accretion rate
distribution, $ M_{\rm KS} = 4.1$ and 3.4, suggesting that the
difference is significant. 

We note that similar conclusions were also reached by \citet{prp02}
and \citet{prp03}. Here one needs to be cautious that, although the
numbers of LMXBs in the Galactic disk with measured orbital periods
and accretion rates have been substantially increased (51 and 45
respectively) since then, this comparison is still severely subject
to small number statistics as well as observational selection
effects, e.g., luminous sources are more likely to be observed.
Since in this work several key procedures involved in the formation
processes of I/LMXBs have been significantly upgraded, the accretion
rate discrepancy, if really exists, strongly suggests that something
is missing or needs to be modified when modeling the evolution of
LMXBs.

A possible mechanism that may help solve the problem is that there
are extra AML mechanisms in the LMXB evolution that are not taken
into account. The calculated accretion rate distribution is peaked
around $\lesssim 10^{-10}\,M_{\sun}$ yr$^{-1}$, implying that the
mass transfer is driven by GR. There is similar situation in CVs, in
which AML is thought to be dominated by MB and GR above and below
the period gap, respectively. However, modeling CV evolution
indicates that the AML rate  below the period gap is $2.47(\pm
0.22)$ times the GR rate \citep{kbp11}, suggesting the existence of
some other AML mechanisms. \citet{sl12a} considered several kinds of
AML mechanisms including isotropic wind from NSs, outflows from the
inner and outer Langrangian points, and the formation of a
circumbinary disk. They found that only outflow from the outer
Langrangian point (or a circumbinary disk) can account for the extra
AML rate, provided that $15-45\%$ of the transferred mass
is lost from the binary. These consequential AML mechanisms
can not only substantially enhance the mass accretion rates, but
also influence the value of the bifurcation periods
\citep[e.g.,][]{ml09}.

Another possible origin of the accretion rate discrepancy is that
the calculated secular mass transfer rate may not be identical to
the accretion rate derived based on short time (no more than $\sim
40$ years) observations. For example, X-ray irradiation of the donor
star may driven mass transfer cycles in LMXBs \citep{p91,br04,b15}.
In this picture, once mass transfer begins, the accretion luminosity
enhances the irradiation flux on the donor. Thermal expansion of the
donor subsequently triggers an increasing mass transfer rate (this
stage is denoted as high state).
The donor evolves to a new thermal equilibrium state on
Kelvin-Helmholtz timescale. When the donor nearly reaches the
equilibrium state, its expansion rate decreases, then the
mass transfer rate drops, and so does the irradiation flux. The
donor starts to contract, and its intrinsic luminosity increases.
This reduces the mass transfer to a quite low rate (denoted as low
state). Hence the LMXB undergoes cyclic mass transfer alternating
between high and low states. Although spending most of their
evolutionary time in low state, they are more likely to be detected
in high state, in which the mass transfer rates can be significantly
larger than the average ones.

In both cases mentioned above the X-ray lifetime is shortened, which
can increase the birthrates of LMXBs derived from the observed
number in the Galactic disk. However, the detailed processes related
to either extra AML loss  or cyclic mass transfer are rather
uncertain and depend on poorly known input parameters.
Numerical calculations show that irradiation in LMXBs may
boost the mass transfer rates by tens of times \citep{br04,b15}.
Furthermore, \citet{r08} argued that the mass transfer cycles are
not likely to work for transient sources with an unstable disk,
because irradiation cannot be sustained in a long quiescent stage.
To illustrate the effect of irradiation in a simplified way, we
arbitrarily increase the mass transfer rates of persistent LMXBs by
a factor of $\rm \Gamma $, which is randomly distributed between 1
and 30. The corresponding evolutionary time is then reduced by the
same factor. The resultant distribution is plotted in Fig.~11, and
we see that it seems to better match the observational data,
especially for the luminous sources. The values of $ M_{\rm KS}$
decreases to be 3.8 and 2.9 correspondingly.

\subsection{Orbital period distribution of BMSPs}

Similar as in Paper I, we follow the evolution of LMXBs to the
formation of binary MSPs, and plot their orbital period
distributions with the black lines in Fig.~11. In most cases the
mass transfer terminates when the donor star loses its hydrogen-rich
envelope and becomes a He or CO WD. For converging LMXBs, we do not
follow the evolution when the systems nearly reach the minimum
periods. They might evolve to be
black widow systems \citep[e.g.,][]{c13,b14}. At the end of each
evolutionary sequence, we know the final orbital period and the
secondary mass. By summing the calculated sequences, we obtain  the
orbital period distributions of BMSPs.  We assume that the lifetimes
of all BMSPs are  $10^{9} $ yr, so their numbers in each
bin of the orbital period depends on their birthrates.

In Fig.~12 we also plot the observed orbital period distributions of
binary pulsars in the red lines. The upper and lower panels
correspond to the initial NS mass of 1.0$ M_{\odot}$  and 1.8$
M_{\odot}$, respectively. Depending on the amount of the accreted
mass, the NSs may be fully or partially recycled. It is well known
that accretion of $\sim 0.1\,M_{\sun}$ material is sufficient to
spin up a NS to millisecond period \citep[e.g.,][]{t12}, so
we distinguish BMSPs from other partially recycled pulsars
using this criterion. In the left panel we show the
theoretical and observational histograms for both fully and
partially recycled NSs, that is, those with any accreted
mass $\Delta M_{\rm NS}\geq 0$, and those with measured spin periods
$ P_{\rm s} \leq$ 1 s, respectively. In the right panel we are
confined with BMSPs, that is, those with $\Delta M_{\rm NS}\geq 0.1
M _{\odot}$, and those with $P_{\rm s} \leq$ 10 ms \citep[data from
the ATNF pulsar
catalogue\footnote{http://www.atnf.csiro.au/research/pulsar/psrcat/};][]{m05}.

The orbital periods of Galactic binary pulsars extend to $ \sim $
1000 days, in agreement with our calculated results with $\Delta
M_{\rm NS}\geq 0$, while those of BMSPs tend to have
systematically small values. The known BMSPs have orbital periods
up to 200 days, longer than predicted if $M_{\rm
NS}=1\,M_{\sun}$ but likely compatible with that if $M_{\rm
NS}=1.8\,M_{\sun}$. However, our calculated distributions show a
shortage of BMSPs with $ P_{\rm orb}\sim0.1-10$ day, which conflicts
with observations \citep[see also][]{prp03}. Figure 13
presents the CDFs for the orbital periods of binary pulsars with
$P_{\rm orb}>0.1$ day. Similar as in Fig.~12, the left and right
panels correspond to all binary pulsars and BMSPs, respectively. The
line styles are same as in Fig.~10. There are 115 and 77 binary
pulsars in the Galactic disk with $ P_{\rm s} < 1$ s and $ P_{\rm s}
< 10$ ms in the ATNF pulsar catalogue, respectively. At a
significance level $ \alpha =0.05$, $ M_{\rm KS}^{\rm cr} \simeq
1.4$. In the left panel, $ M_{\rm KS} \simeq 2.5 $ and 3.9 for $
M_{NS}=1.0 M_{\odot}$ and $1.8 M_{\odot}$, respectively.  In the
right panel, $ M_{\rm KS} \simeq 1.7$ and 2.1 for $ M_{NS}=1.0
M_{\odot}$ and $1.8 M_{\odot}$, respectively. In both cases the $
M_{\rm KS} $ values are larger than  $ M_{\rm KS}^{\rm cr}$  at
$P_{\rm orb}\sim 0.1-10$ days.

The rarity of the predicted narrow BMSPs is likely to be related to
the accretion rate discrepancy discussed before, as the orbital
period evolution largely depends on the mass transfer process. It is
caused by the fact that in our (and almost all previous)
calculations RLOF is mainly driven by nuclear evolution of the donor
or MB, so that the bifurcation period is around $0.3-1$ day
\citep{ps88,ps89,ml09}. For binaries with initial orbital
periods around it, mass transfer will finally lead to either wider
or narrower orbits. Only with the initial parameters in a very small
parameter space can LMXBs evolve into this orbital period range. It
seems that neither changing the MB law nor considering the feedback
of X-ray irradiation can solve this problem in a satisfactory manner
\citep[see also][for a relevant discussion]{itl14}. Similar problems
also exist for the formation of UCXBs. It was found that with the
standard MB law, only a very small range of initial parameters are
allowed for the binaries to evolve to UCXBs within Hubble time
\citep{prp02,vs05a}; if a saturated MB law is adopted, no UCXBs can
be formed at all \citep{vs05b}. This makes it impossible to account
for the relatively large number of observed UCXBs. All these facts
point to some missing (or misunderstood) AML mechanisms in the LMXB
evolution, which will be investigated in a future work.

\section{Summary}

In this work, we incorporate detailed evolutionary calculations with
BPS to explore the formation and evolutionary tracks of I/LMXBs in
the Galaxy. By this way, we can trace the evolution of the I/LMXBs
from their incipient stage to the descendant binary pulsars. Here we
summarize the results as follows.

1. The incipient I/LMXBs contain NSs formed by both core-collapse
and electron-capture SNe. Their orbital periods are distributed in
the range of $\sim 0.2-100 $ days and $\sim 0.2-1000 $ days,
respectively. The total birthrate of Galactic I/LMXB population is
$\sim 9\times10^{-6}- 3.4\times10^{-5}\,\rm yr^{-1}$ with a star
formation rate of $5M_{\sun}$\,yr$^{-1}$.

2. Since the observability of I/LMXBs depends on the X-ray lifetimes
and the stability of the accretion disks, Galactic LMXBs tend to be
compact binaries with orbital periods $ \lesssim $ 1 day and donor
masses $ M_{\rm d}\lesssim 0.3 M_{\odot} $. These features are
consistent with those of observed LMXBs. However, the average mass
transfer rate  $ \lesssim 10^{-10} M_{\odot} $\,yr$^{-1}$
for persistent X-ray binaries is considerably lower than
observed.

3. The orbital periods of BMSPs are predicted to extend from less
than 1 day to hundreds of days, covering the majority of observed
BMSPs. However, the number of BMSPs with $P_{\rm orb}\sim 0.1-10$
day is severely underestimated.

4. The  conflicts mentioned above
imply that something is missing in the modeling of
LMXBs, which is likely to be related to the (unknown) AML mechanisms.

\begin{acknowledgements}

This work was supported by the Natural Science Foundation of China
under grant numbers 11133001, 11203009 and 11333004, and the
Strategic Priority Research Program of CAS (under grant number
XDB09000000).

\end{acknowledgements}

\clearpage

\begin{figure}

\plottwo{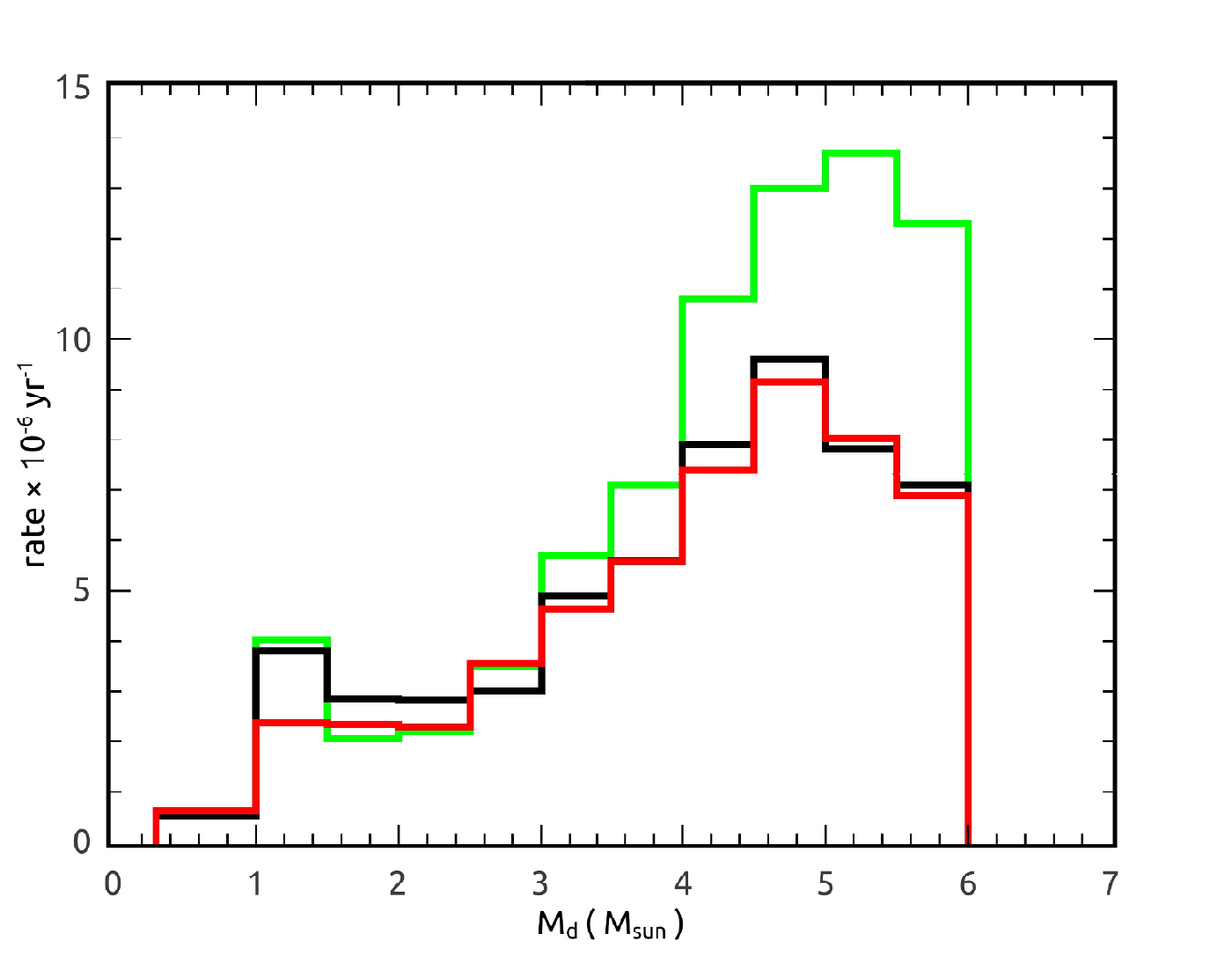}{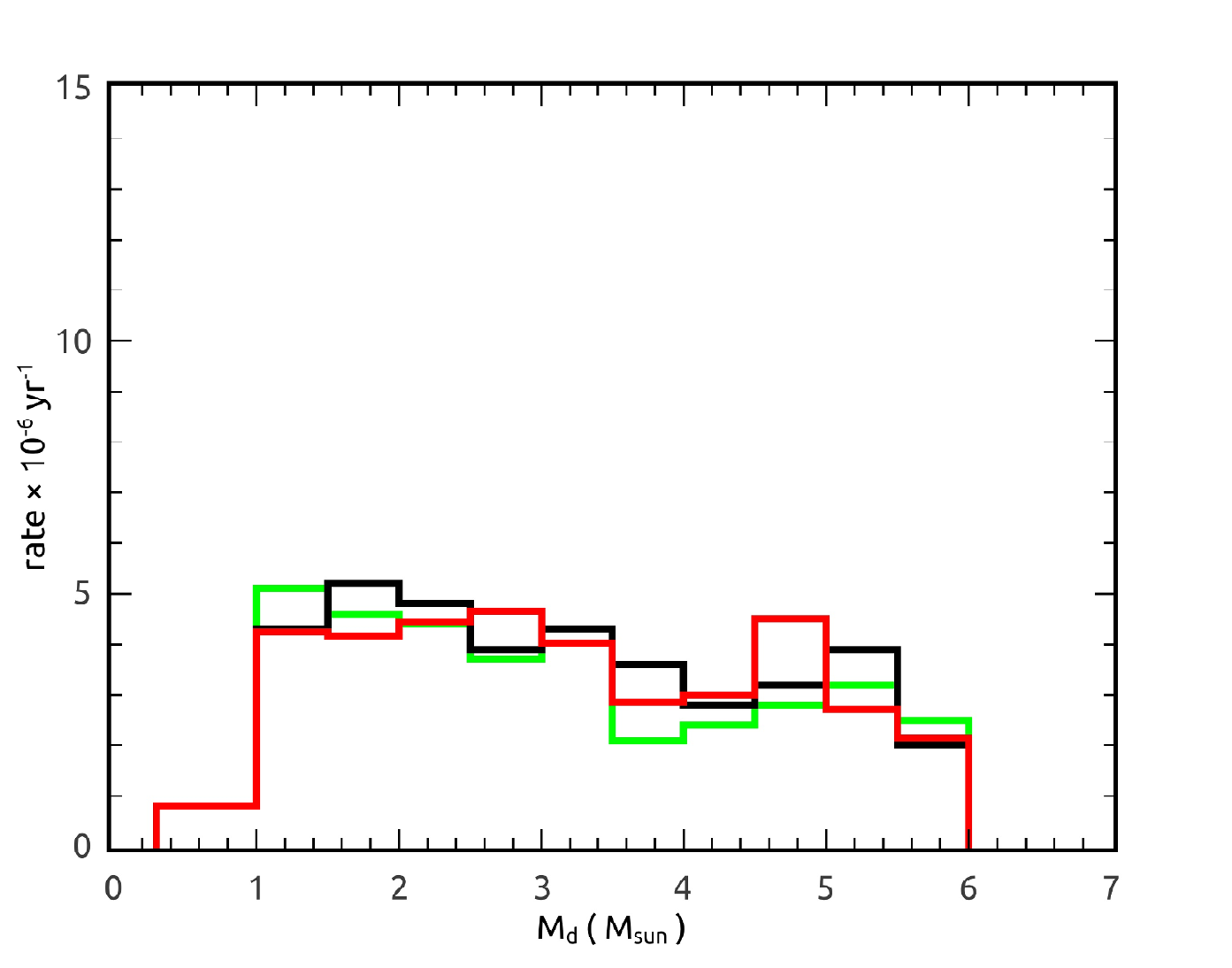}
\caption{ The birthrates of incipient I/LMXBs as a function of the
donor mass $ M_{\rm d}$ derived from the BPS calculations.
Binaries containing NSs descended from core-collapse and electron-capture
SNe are shown in the left and right panels, respectively. The green, red
and black curves represent the results with the $ q_{\rm c} $ values
in Models I, II and III of \citet{sl14}, respectively.
     \label{figure}}

\end{figure}

\begin{figure}

\plottwo{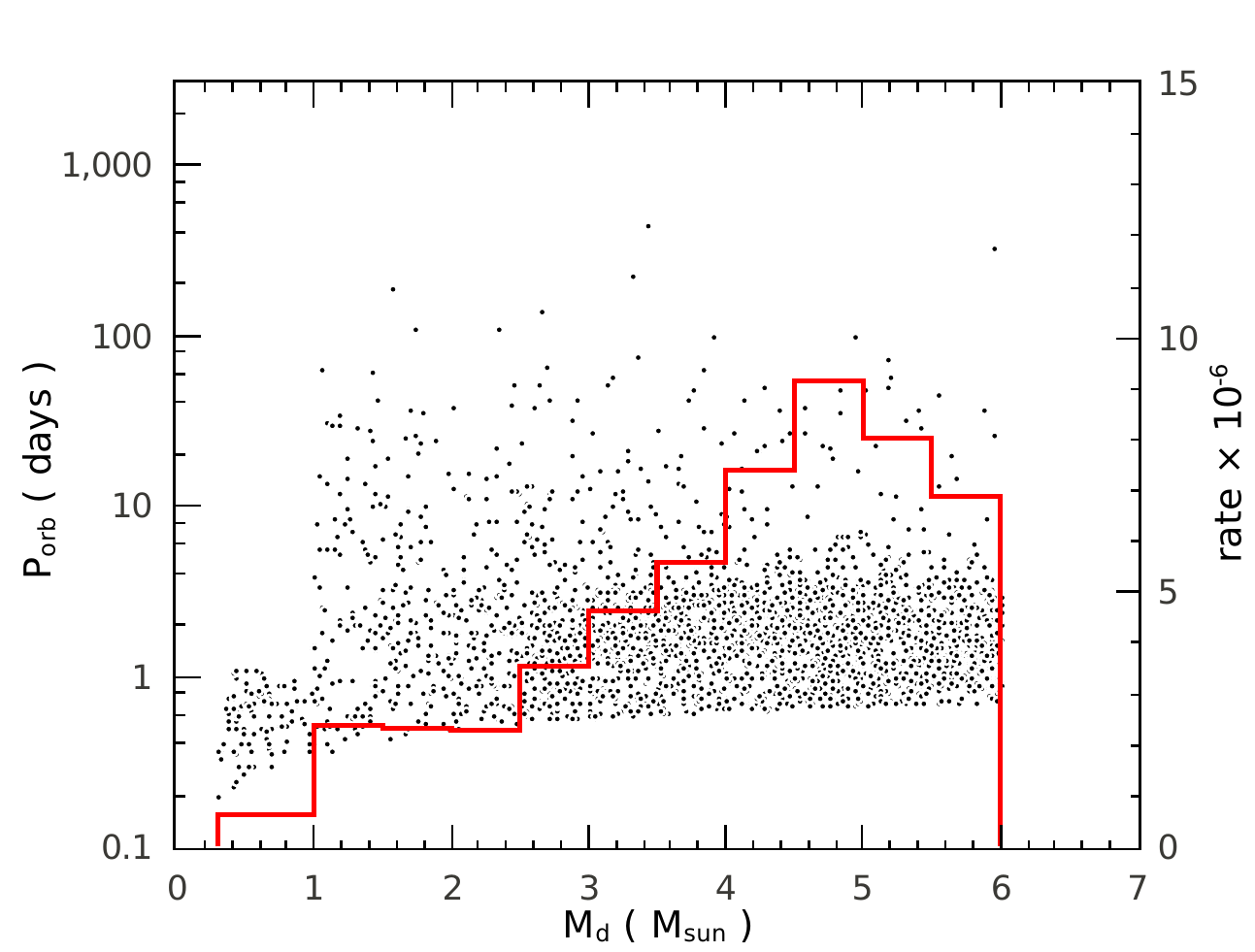}{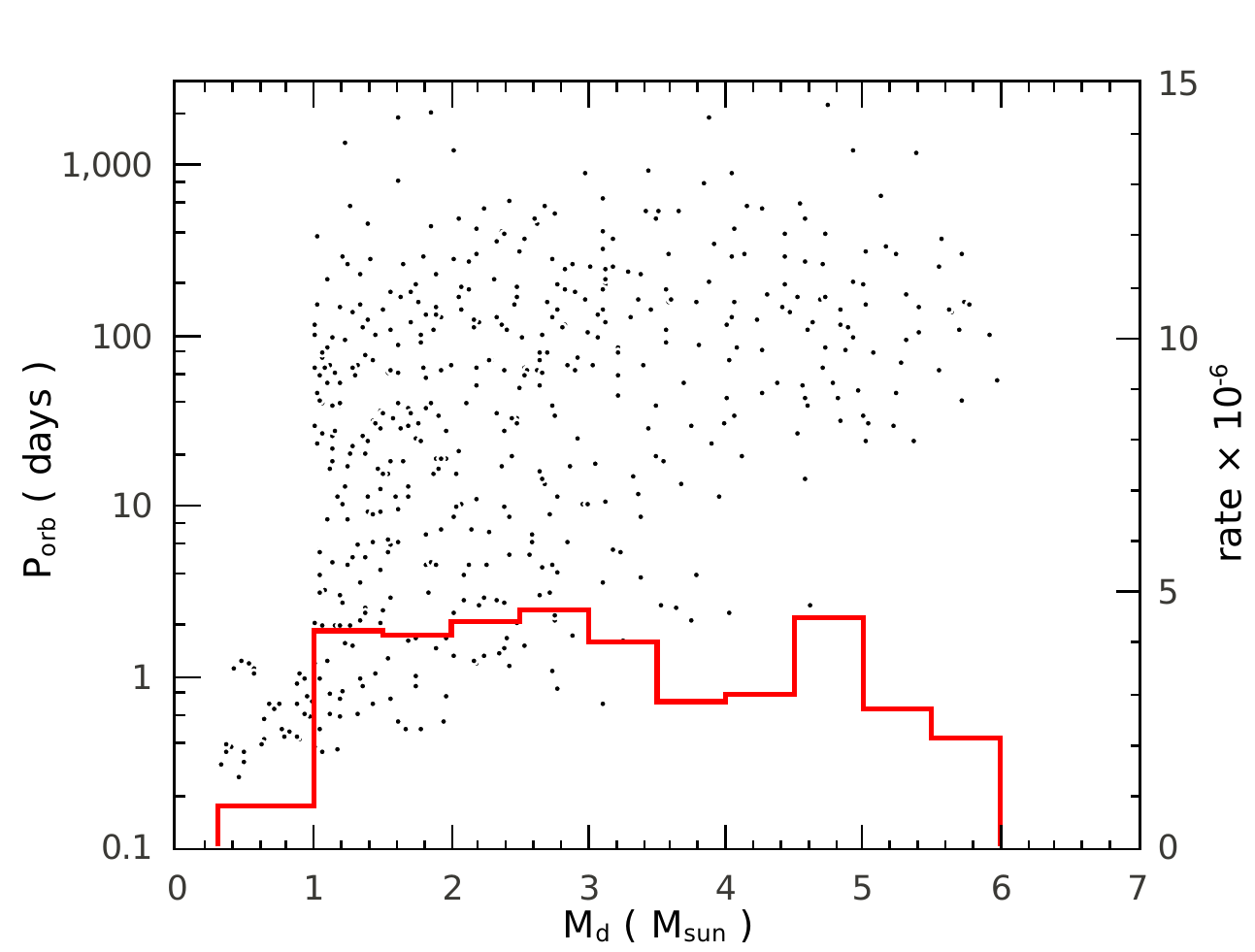}
\caption{ Distributions of incipient I/LMXBs in the plane of initial donor
masses $ M_{\rm d}$ and orbital periods $P_{\rm orb} $.
Binaries containing NSs descended from core-collapse and electron-capture
SNe are shown in the left and right panels, respectively. The red curves
reflect their birthrates (in Model II) as a function of $ M_{\rm d} $.
     \label{figure}}

\end{figure}

\clearpage

\begin{figure}[h,t]
\includegraphics[scale=0.6]{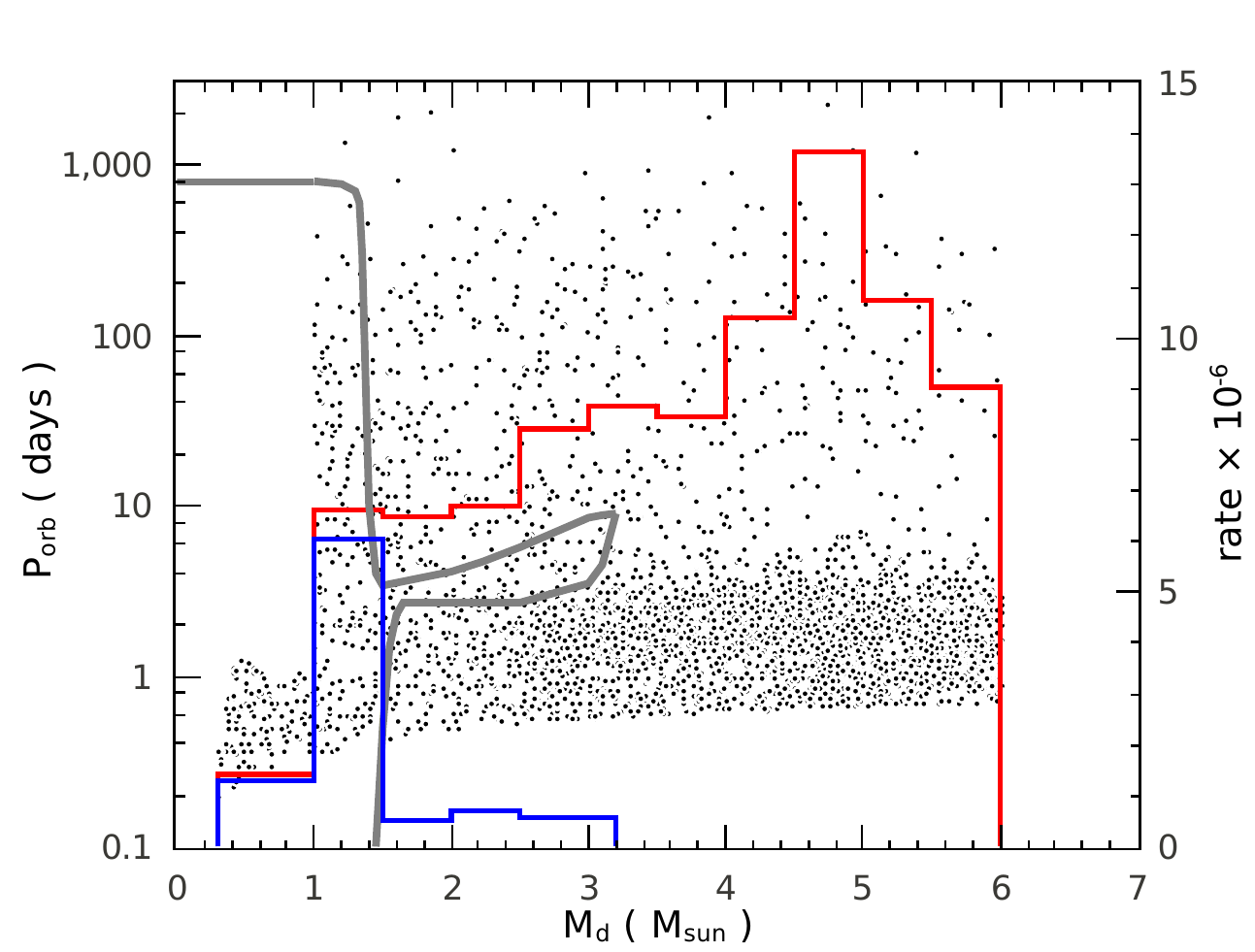}
\includegraphics[scale=0.6]{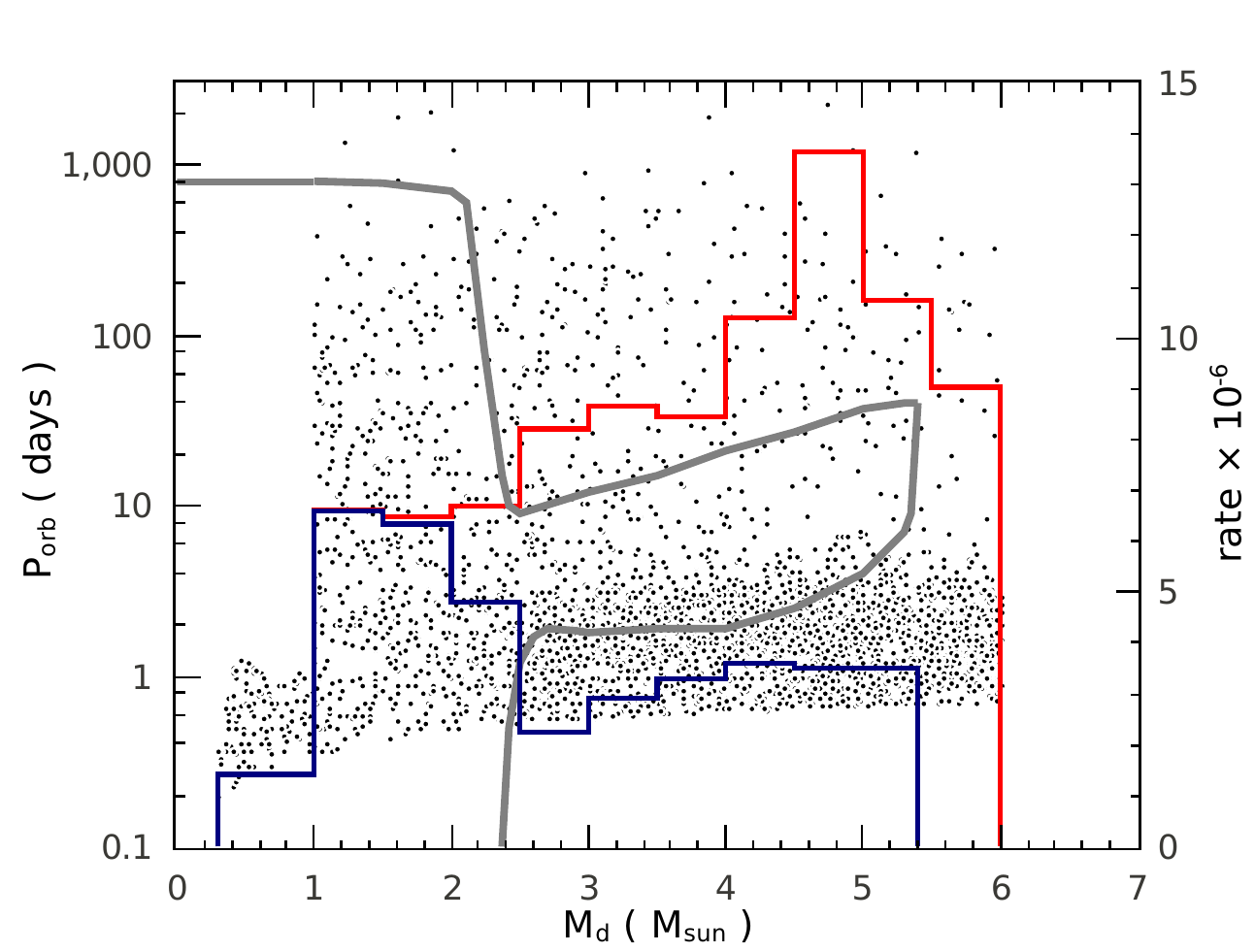}

\caption{Distributions of incipient I/LMXBs (from both core-collapse
and electron-capture SN channels) in the $ M_{\rm d}-P_{\rm orb} $
plane. The red curves reflect their birthrates as a function of $
M_{\rm d} $, same as in Fig.~2. The grey curves confine the
parameter space for stable mass transfer in X-ray binaries with an
initial 1.0$ M_{\odot} $ (left panel) and 1.8$ M_{\odot} $ (right
panel) NS, and the birthrates for binaries within it are shown with
the blue curves.
     \label{figure}}

\end{figure}

\begin{figure}[h,t]
\includegraphics[scale=0.6]{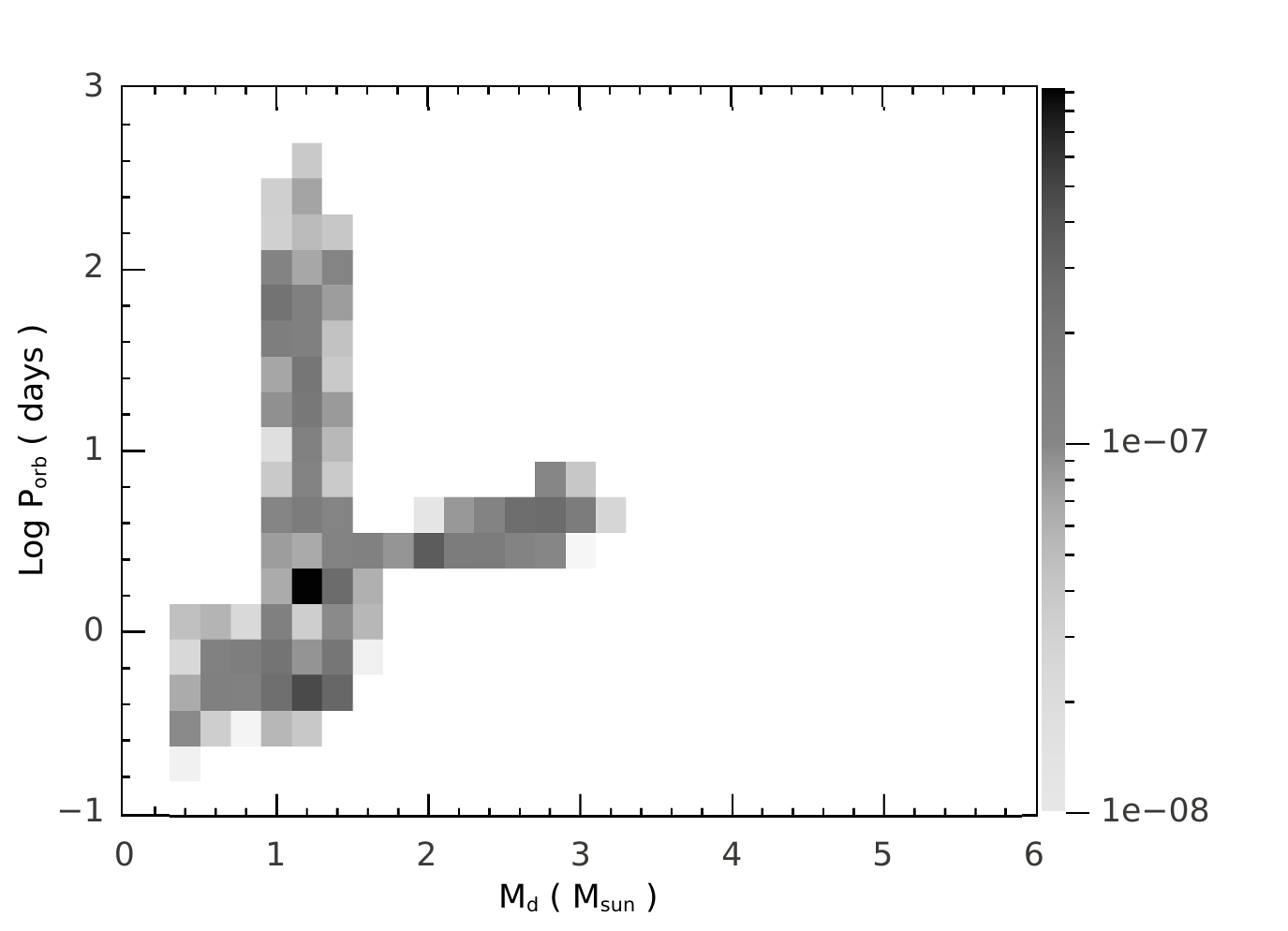}
\includegraphics[scale=0.6]{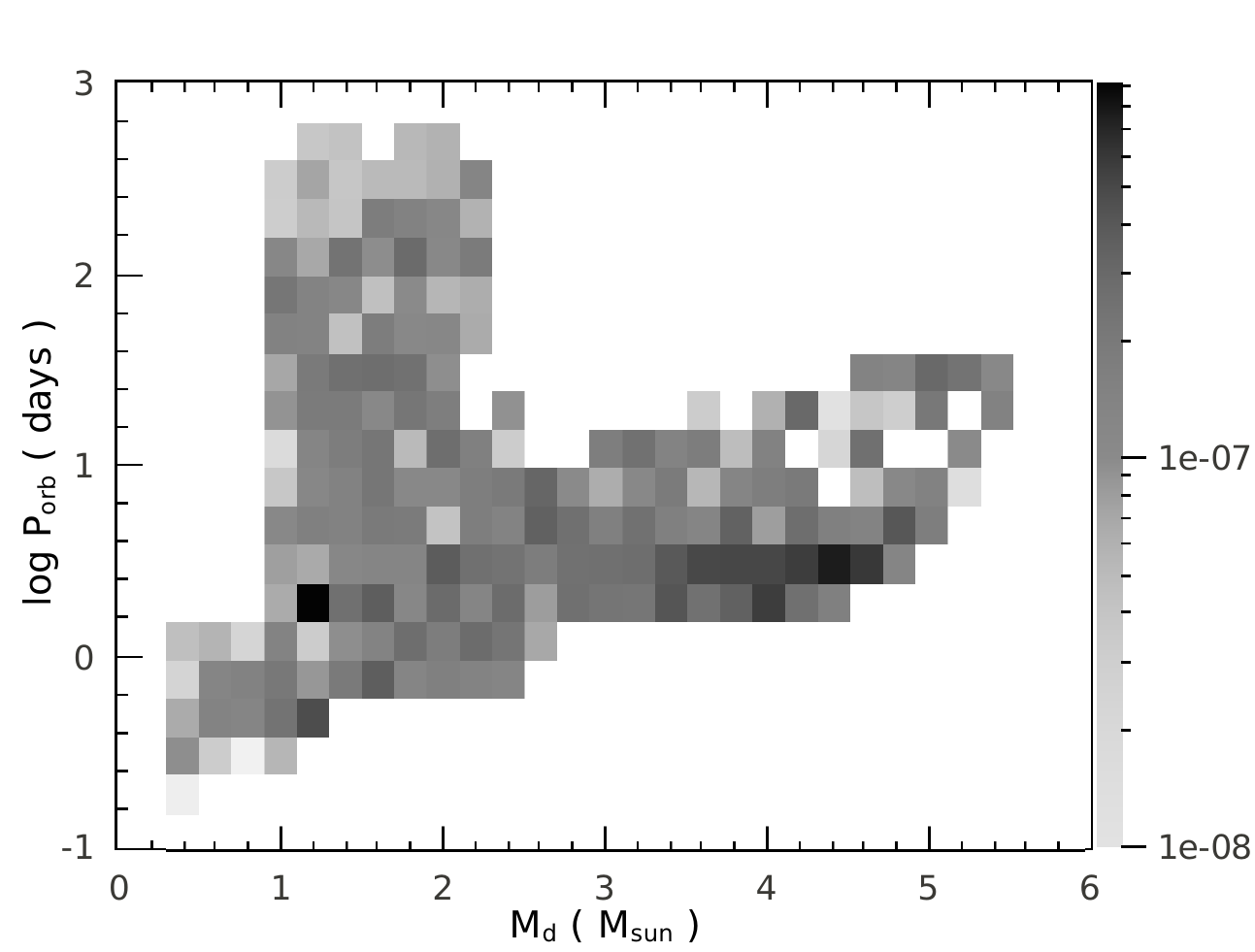}

\caption{Distributions of of incipient I/LMXBs that have stable mass transfer
in the $ M_{\rm d}-P_{\rm orb} $ plane.
 \label{figure}}

\end{figure}

\clearpage

\begin{figure}[h,t]

\centerline{\includegraphics[angle=-90,width=1.0\textwidth]{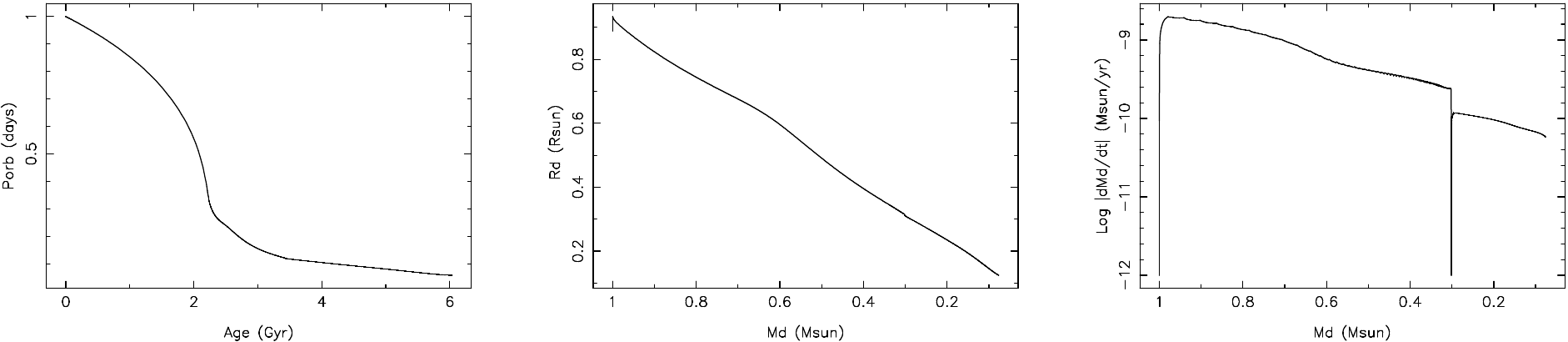}}

\caption{Evolutionary tracks of a LMXB with  initial parameters of
 $ M_{\rm NS} = 1.8 M_{\odot}$, $ M_{\rm d} = 1.0
M_{\odot}$ and $ P_{\rm orb}  = 1.0$ day. The left, middle, and
right panels depict the orbital period $ P_{\rm
orb} $ as a function of age, the  donor radius $ R_{\rm d} $
and the mass transfer rate $ |\dot{M}_{\rm d}| $ as a function of the donor mass
$ M_{\rm d} $, respectively. \label{figure}}

\end{figure}

\begin{figure}[h,t]

\centerline{\includegraphics[angle=-90,width=1.0\textwidth]{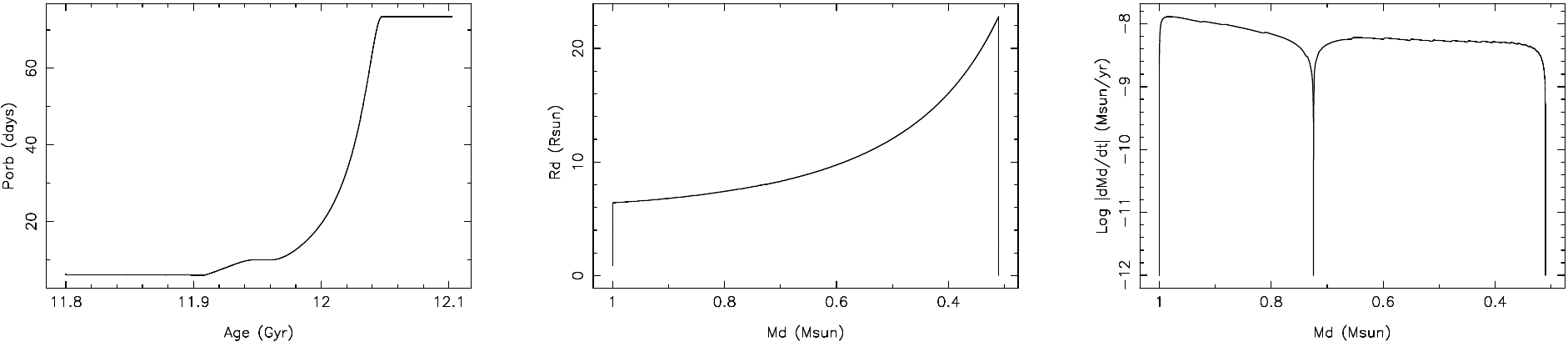}}

\caption{Similar to Fig.~5 but for an initial
 orbital period $P_{\rm orb}=$ 6.3 days.
\label{figure}}

\end{figure}

\clearpage

\begin{figure}[h,t]
\plottwo{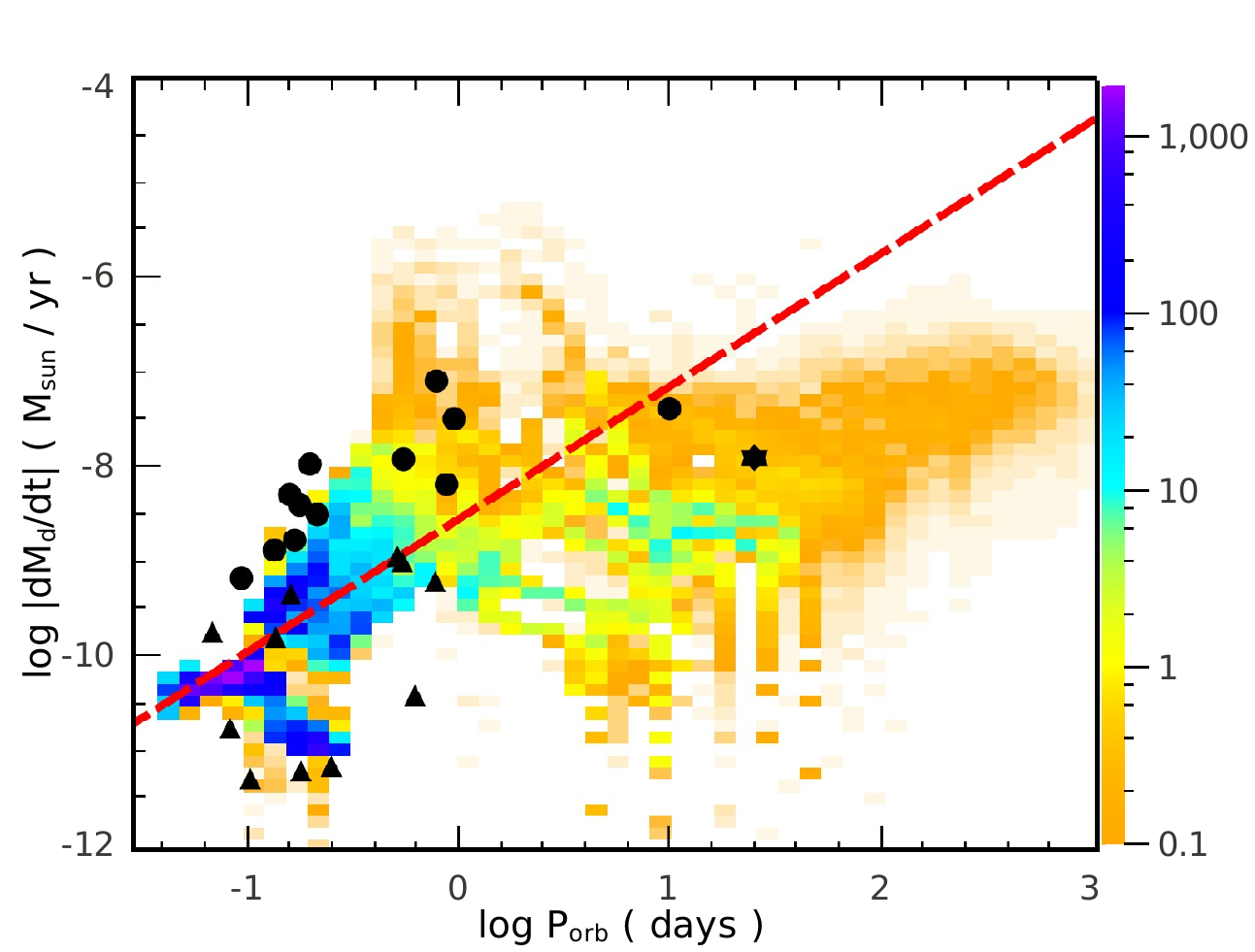}{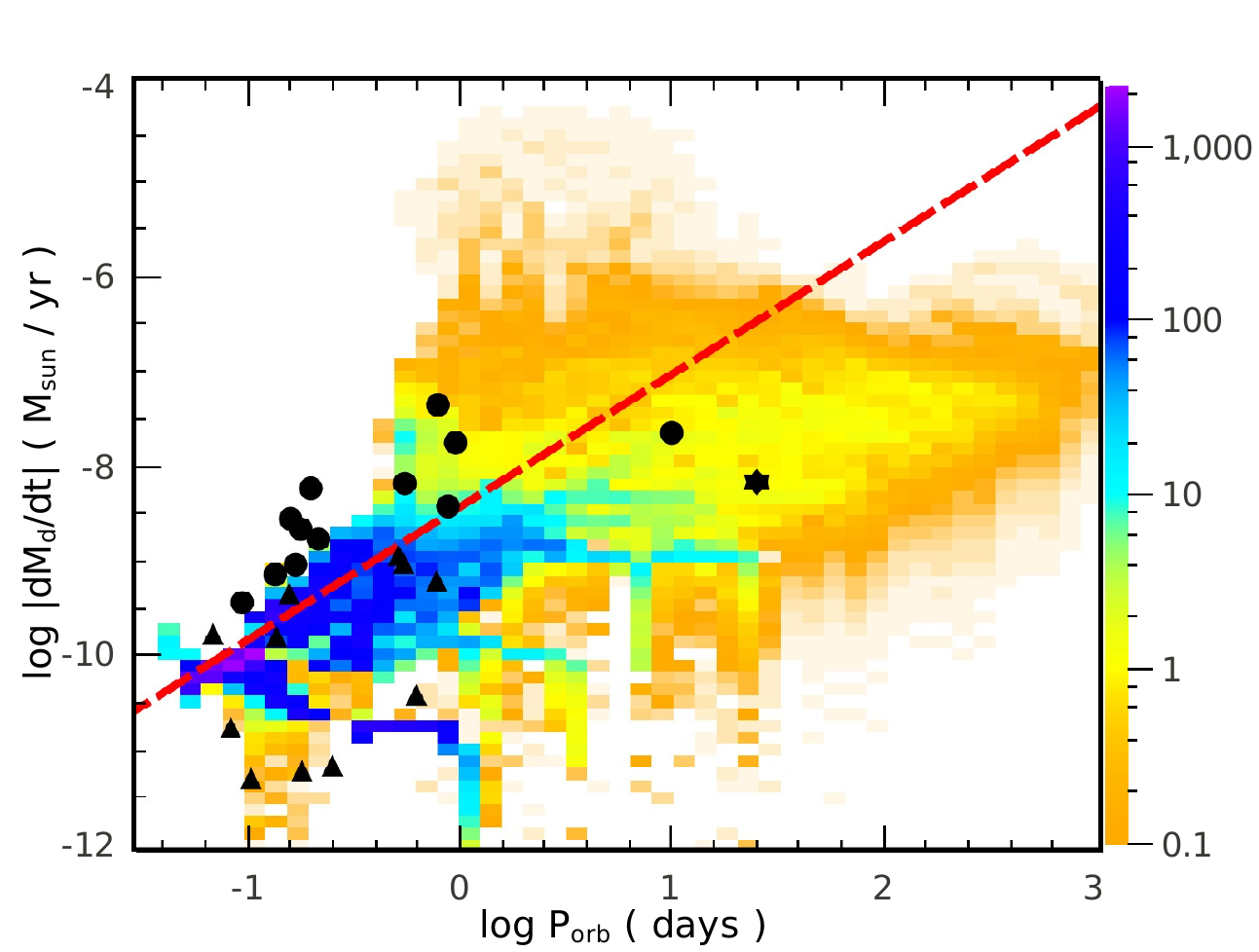}

\caption{The orbital period  and mass transfer rate distributions
for I/LMXBs with a 1.0$ M_{\odot} $ (left panel) and 1.8$ M_{\odot}
$ (right panel) NS. The colors represent the numbers of binaries in
each pixel. The red dash line reflects the critical mass transfer
rate for unstable accretion disks. The filled circles and triangles
represent persistent  and transient LMXBs in the Galactic disk,
respectively. The filled star corresponds to GX 13+1.
  \label{figure}}

\end{figure}

\clearpage

\begin{figure}[h,t]

\plotone{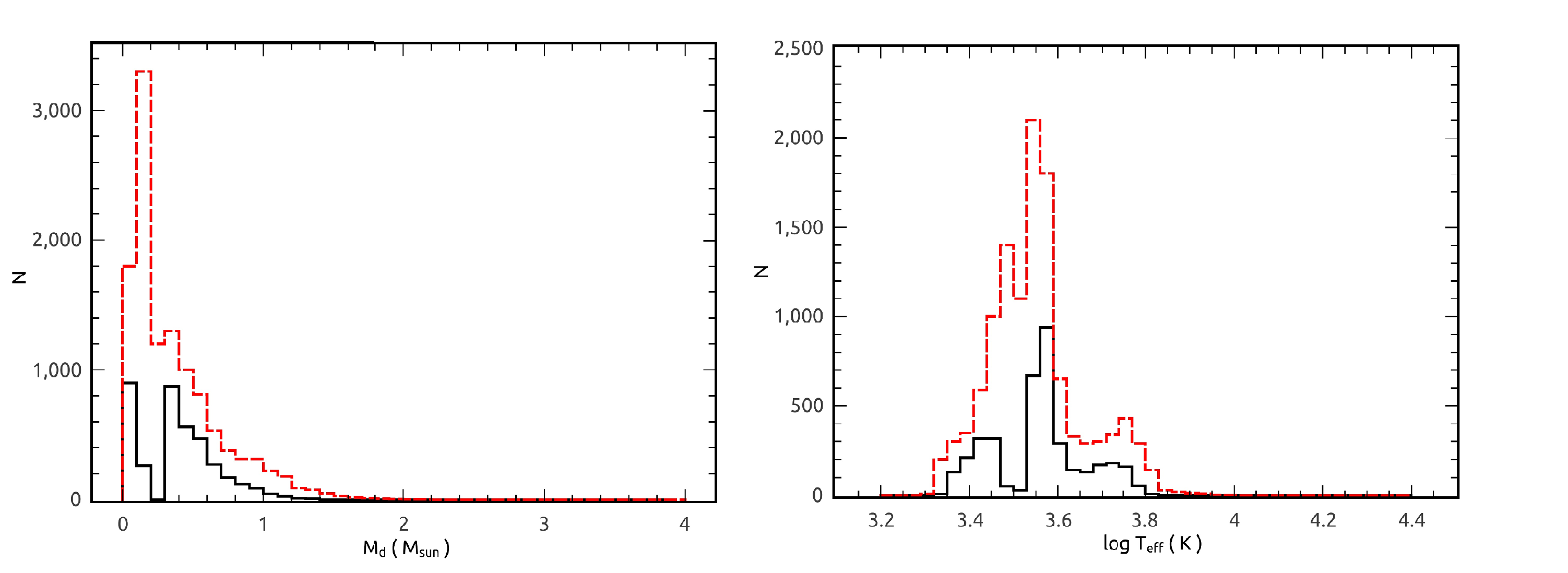}\plotone{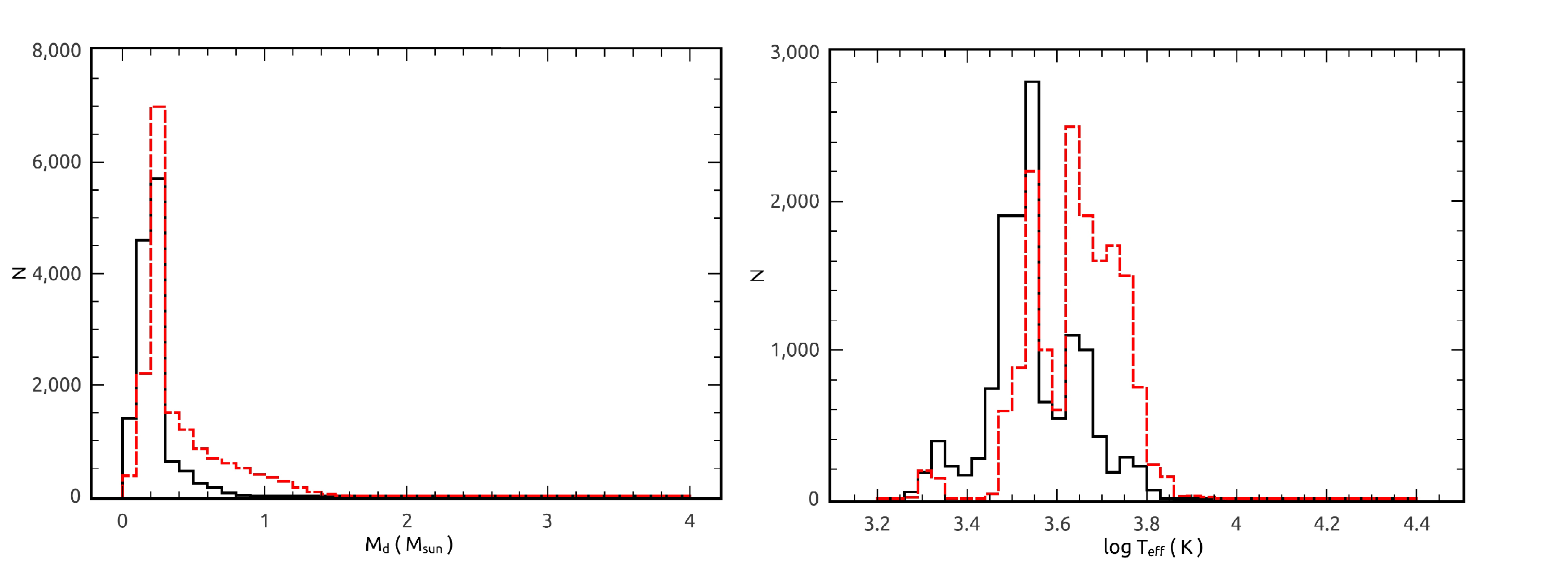}

\caption{The calculated distributions of the donor mass (left panel)
and effective temperature (right panel) in the I/LMXBs. The top and
bottom panels show persistent and transient X-ray binaries,
respectively. The black and red curves correspond to the initial NS
mass of $1.0 M_{\odot} $ and $1.8 M_{\odot}$, respectively.
\label{figure}}

\end{figure}

\begin{figure}[h,t]
\plotone{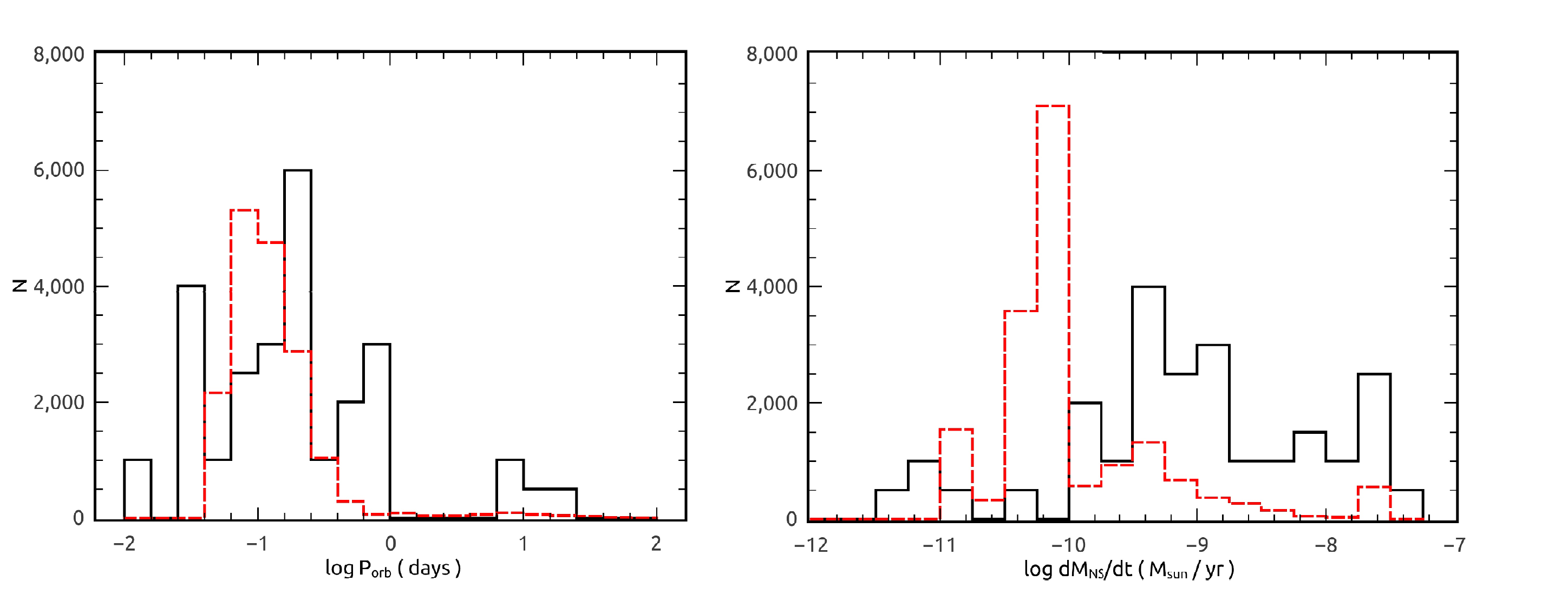}
\plotone{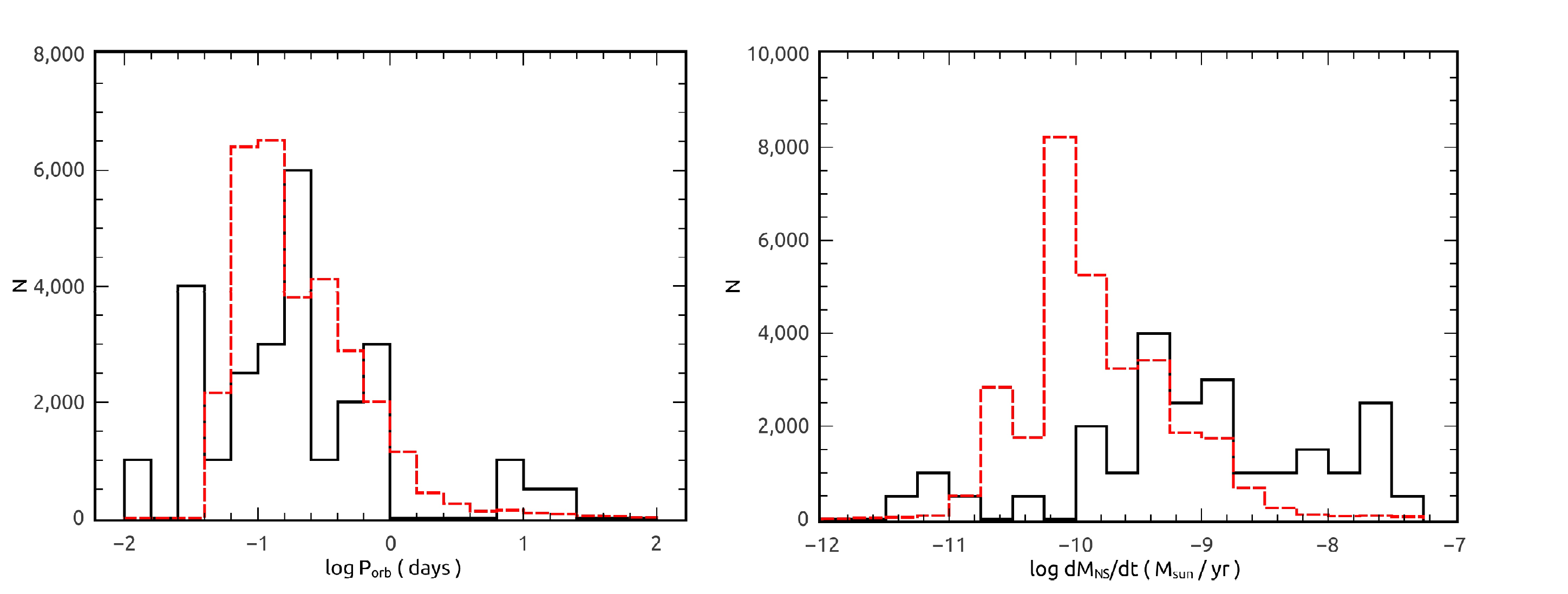}

\caption{Comparison of the calculated orbital period (left panel)
and accretion rate (right panel) distributions of Galactic I/LMXBs
with observations, which are shown with the red and black curves,
respectively. The top and bottom panels correspond to the initial NS
mass of $1.0 M_{\odot} $ and $1.8 M_{\odot}$, respectively.
  \label{figure}}

\end{figure}

\clearpage

\begin{figure}
\includegraphics[scale=0.6]{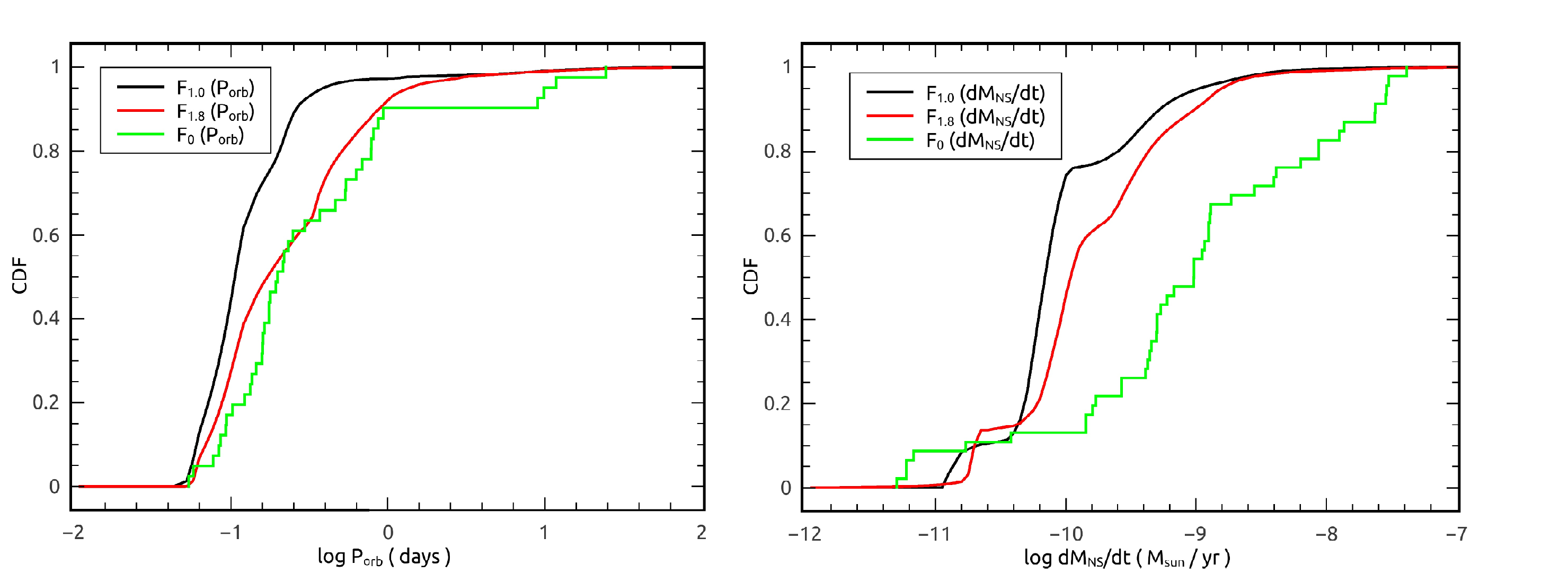}
\caption{Cumulative distribution functions for the orbital period
(left panel) and the accretion rate (right panel) of Galactic
I/LMXBs. The green curve shows the observed distribution function.
The black and red curves correspond to the calculated distribution
functions with a $1.0 M_{\odot} $ and $1.8 M_{\odot} $ NS,
respectively. \label{figure}}
\end{figure}

\clearpage

\begin{figure}[h,t]
\includegraphics[scale=0.6]{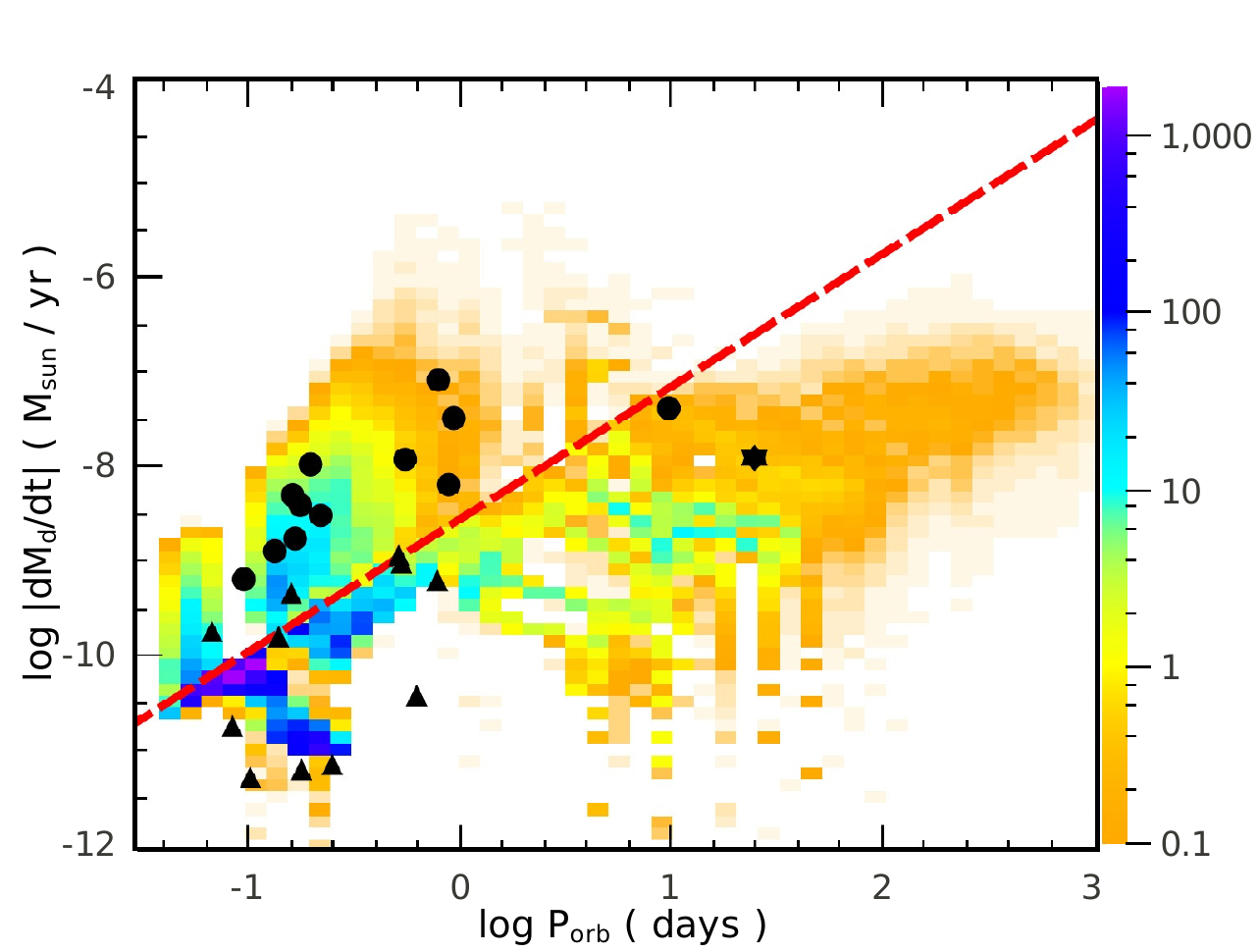}
\includegraphics[scale=0.6]{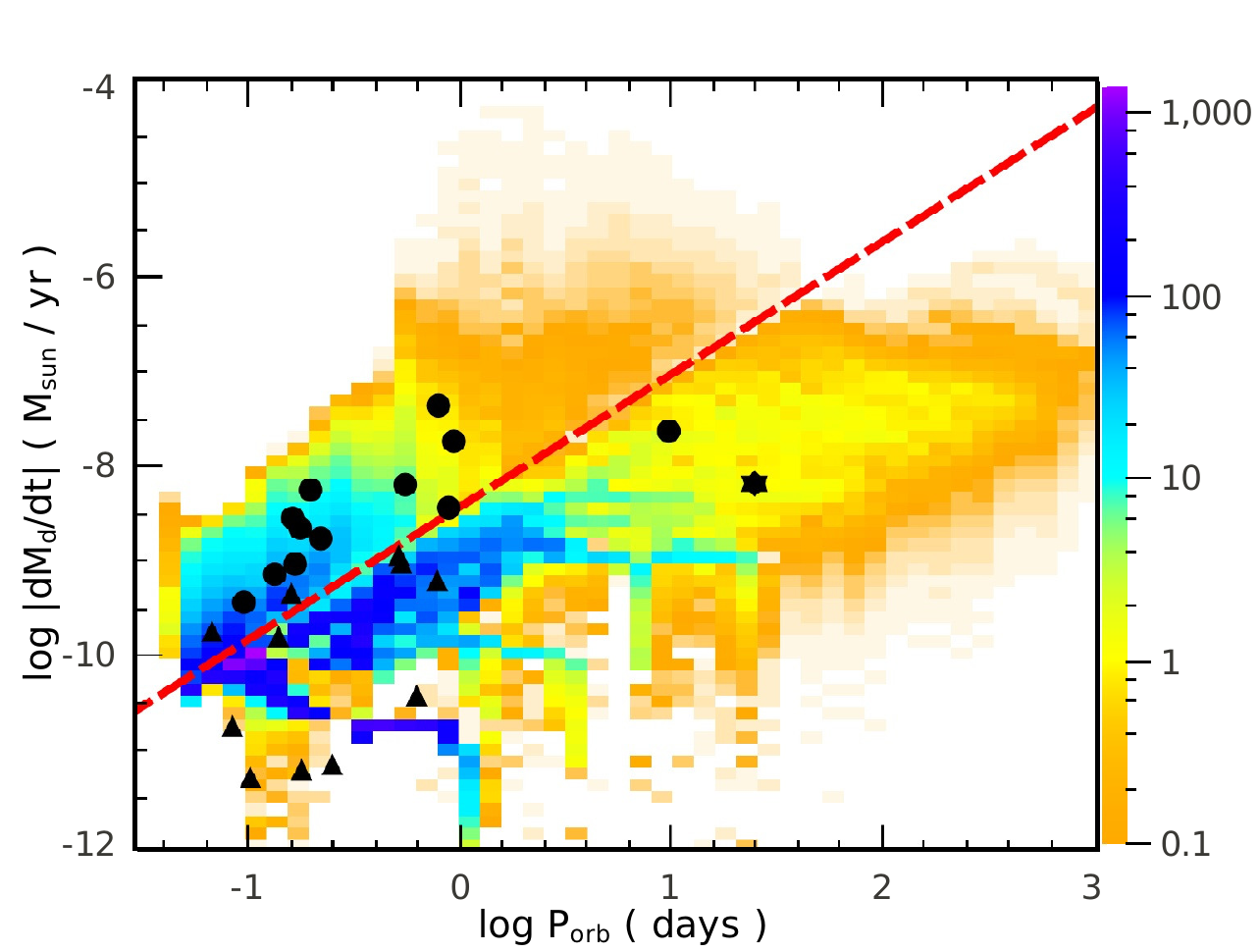}

\caption{Similar as Fig.~7, but with the irradiation-induced mass
transfer cycles taken into account. The mass transfer rates for
persistent sources are arbitrarily increased by a factor between 1
and 30, while their lifetimes are decreased by the same factor.
  \label{figure}}

\end{figure}

\begin{figure}[h,t]
\plotone{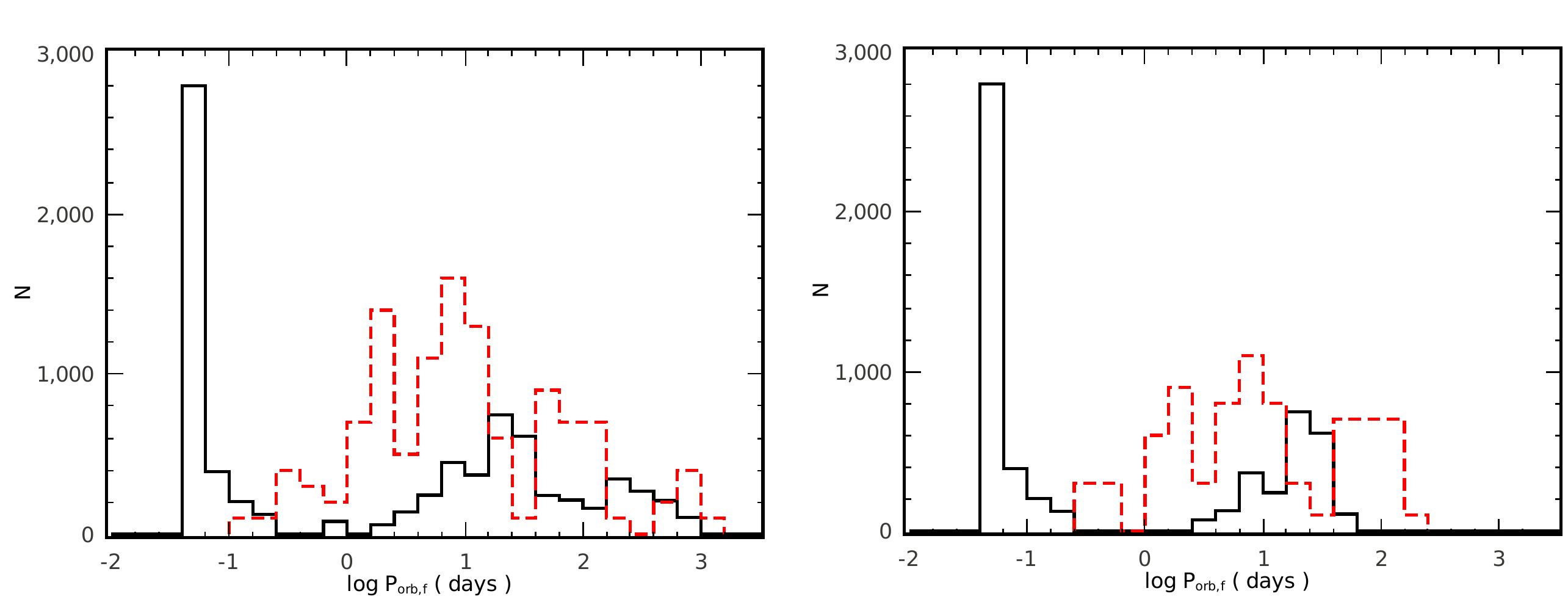} \plotone{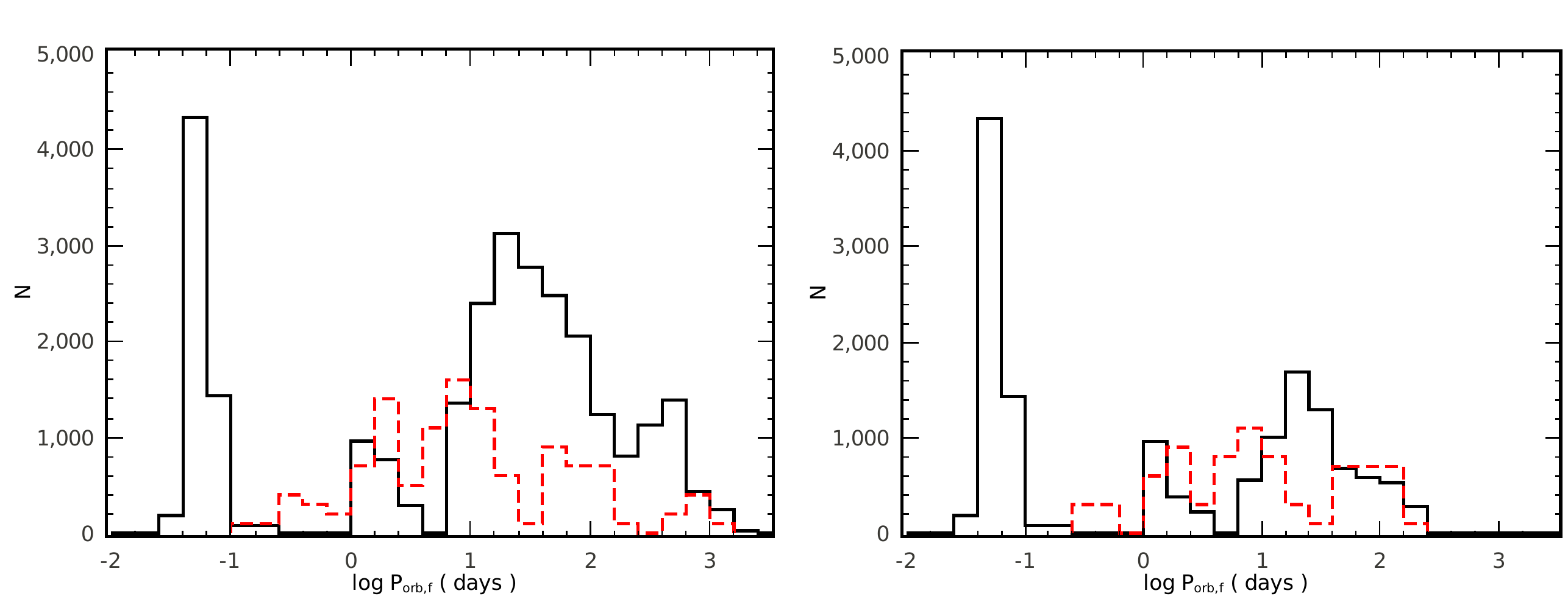}

\caption{Comparison of the calculated and measured orbital period
distributions of binary pulsars, which are shown with the black and
red curves, respectively. The top and bottom panels correspond to
the initial NS mass of $1.0 M_{\odot} $ and $1.8 M_{\odot}$,
respectively. The left panel shows the calculated distributions of
pulsars with any accreted mass $ \Delta M_{\rm NS} > 0$ and
the observed distribution of binary pulsars with spin periods $
P_{\rm s} \leq 1$ s. The right panel shows the calculated
distributions of pulsars with accreted mass $ \Delta M_{\rm NS} \geq
0.1 M_{\odot}$ and the observed distribution of binary pulsars with
spin periods $ P_{\rm s} \leq 10$ ms.
  \label{figure}}

\end{figure}

\begin{figure}
\includegraphics[scale=0.6]{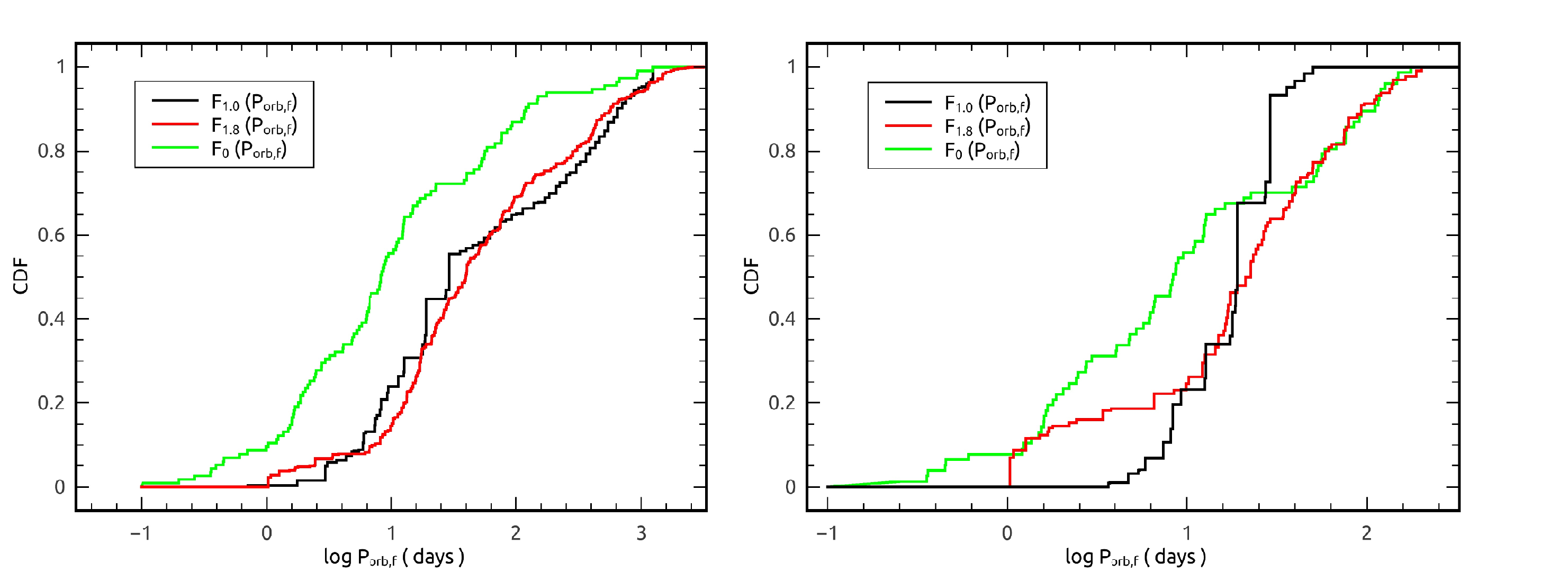}
\caption{Cumulative distribution functions for the  orbital periods
of binary pulsars (left panel) and BMSPs (right panel). The line
styles are same as in Fig.~11. \label{figure}}

\end{figure}

\end{document}